\documentclass[prb,twocolumn,showpacs,superscriptaddress]{revtex4-1}

\usepackage{graphicx}
\usepackage{amsmath,amssymb,bm}

\usepackage{color}

\begin{document}


\title{
$1/(N-1)$ expansion for a finite $U$ 
Anderson model away from half-filling
}


\author{Akira Oguri}
\affiliation{%
Department of Physics, Osaka City University, Sumiyoshi-ku, 
Osaka, Japan
}


\date{\today}


\begin{abstract}
We apply recently developed $1/(N-1)$ expansion 
to a particle-hole asymmetric 
SU($N$) Anderson model with  finite Coulomb interaction $U$.
To leading order in $1/(N-1)$   
it describes the Hartree-Fock random phase approximation (HF-RPA), 
and the higher-order corrections describe systematically 
the fluctuations beyond the HF-RPA.
We show that the next-leading order results of  
the renormalized parameters for the local Fermi-liquid  state 
agree closely with the numerical 
renormalization group results at $N=4$.  
It ensures the reliability of the next-leading order results for $N \geq 4$,
 and we examine the $N$ dependence of the local Fermi-liquid parameters. 
Our expansion scheme uses the standard Feynman diagrams,  
and has wide potential applications.
\end{abstract}

\pacs{72.15.Qm, 73.63.Kv, 75.20.Hr}


\maketitle

\section{Introduction}
\label{sec:introduction}

Exploring the ground state and 
the low-energy excitations 
of strongly correlated electron systems is 
one of the fundamental subjects of quantum many-body theory. 
The low-lying energy states determine 
not only  equilibrium properties but 
nonequilibrium transport 
deviating from the linear response.
For instance,  in quantum dots 
the universal Kondo behavior 
of the steady current\cite{KNG,ao2001,FujiiUeda,HBA} 
and shot noise\cite{GogolinKomnik,Golub,Sela2009} 
 at small bias voltages can be described in terms of 
the quasi-particles of a local 
 Fermi liquid. 

The universal behavior has been studied 
in more general situations with the orbital 
degeneracy.\cite{Mora2009,Sakano,SakanoFCS,AoSakanoFujii}
The essential feature of the universality 
in this case may be deduced from the low-energy states of 
the $N$-fold degenerate Anderson impurity. 
The exact numerical renormalization group (NRG) approach\cite{KWW} 
was successfully applied to this model 
 for small degeneracies  $N \leq 4$,\cite{Izumida,Choi,Anders,Nishikawa2,Sakano} which for $N=2$ corresponds to the spin degeneracy.
However, 
alternative approaches are required 
for large degeneracies $N\geq 4$.  

The conventional $1/N$ expansion, or non-crossing approximation (NCA),  
is one of such approaches.\cite{Coleman,Bickers} 
This and related approximations are based on the perturbation theory 
with respect to the hybridization matrix element $V$ 
between the impurity and conduction band.
They treat mainly the electron filling 
 less than one electron in $N$ orbitals,  
assuming an infinite Coulomb interaction $U$. 
Extensions of the methods of this category 
 to the case of finite $U$ have been developed, 
also on the basis of the perturbation theory in $V$, 
using the slave-boson or the resolvent technique.\cite{Haule,OtsukiKuramoto}
However, in order to explore a wide range of 
the electron fillings and of the other parameters, 
it is still necessary to develop different 
approaches applicable to the low-energy excitations.

In a previous paper, 
we proposed  a completely different approach,
using a scaling that takes $u =(N-1)U$ as an independent variable 
instead of the Coulomb interaction $U$ alone.\cite{AoSakanoFujii}
The factor $N-1$ appears naturally 
as it represents the number of interacting orbitals, 
excluding the one prohibited by the Pauli principle. 
With this scaling the perturbation series in $U$  
can be reorganized as an expansion in powers of $1/(N-1)$.
 This can be carried out 
by using a standard Feynman diagrammatic 
technique for the fermions with two-body interactions.
Specifically, diagrammatic classification, 
 similar to the one for the $N$-component $\varphi^4$ model 
for critical phenomena,\cite{WilsonKogut} 
is applicable with an extension that    
takes into account the fermionic anticommutation relation. 
We calculated the renormalization parameters for the quasi-particles  
 up to order $1/(N-1)^2$ at half-filling, and confirmed that 
the results agree closely with the NRG results 
for $N=4$. It enabled us to clarify the $N$ dependence of 
the nonequilibrium Kondo behavior for $N> 4$.\cite{AoSakanoFujii}

In the present paper, we apply this approach 
to away from half-filling, and examine the ground-state properties.
To this end, we choose the unperturbed Green's function such that 
it includes the level shift due to 
the $\omega=0$ component of the self-energy, 
which at zeroth order is given 
by the Hartree-Fock (HF) approximation.
The first order contributions in the $1/(N-1)$ expansion 
consists of the vertex and self-energy corrections, 
which are equivalent 
 to those of the random phase approximation (RPA).\cite{SchmittAnders}
The higher order contributions  
systematically capture the correlation effects 
beyond the HF-RPA. 
We show that the order $1/(N-1)^2$ 
results of the renormalized parameters 
agree   well to those of 
the NRG also in the particle-hole asymmetric case.
It ensures an early convergence of the $1/(N-1)$ expansion 
for $N \geq 4$. 
With this approach, we also examine the $N$ dependence 
of the local Fermi-liquid parameters.
Our expansion procedure, using the standard Feynman diagrams, 
is quite general. It can be applied to the Keldysh formalism, 
and to lattice systems such as the Hubbard model.

This paper is organized as follows. We describe the model
and the basic idea of the scaling of the Coulomb interaction $U$ 
in Sec.\ \ref{sec:formulation}. 
The diagrammatic classification 
for the leading and next-leading order terms in 
 $1/(N-1)$ expansion is explained in Sec.\ \ref{sec:1_(N-1)_expansion}.
The numerical results for the renormalized parameters are 
presented in Sec.\ \ref{sec:results}. 
A brief summary is given in Sec.\ \ref{sec:summary}.

\section{Model and Formulation}
 \label{sec:formulation}

We consider the SU($N$) impurity Anderson model  
with finite interaction $U$,  given by  
$\,{\cal H} \,= \, {\cal H}_{d} + {\cal H}_{c}  
 + {\cal H}_V + {\cal H}_U$, 
\begin{align}
& 
\!
{\cal H}_{d} =    
\sum_{m=1}^N 
 E_d^{}\,n_{dm}^{} , \quad \ \ 
{\cal H}_{c} = \sum_{m=1}^N 
\int_{-D}^D  \! d\epsilon\,  \epsilon\, 
 c^{\dagger}_{\epsilon  m} c_{\epsilon  m}^{},
\\
&
\!
{\cal H}_U 
=    
\sum_{m \neq m'}  
\! 
 \frac{U}{2} \,
\Bigl[  n_{dm}^{} -\langle n_{dm}\rangle \Bigr]
\Bigl[  n_{dm'}^{} -\langle n_{dm'}\rangle \Bigr],
\label{eq:H_U} \\
 &
\! 
\mathcal{H}_V  =  
\sum_{m=1}^N 
 V \! \left(
d_{m}^{\dagger} \psi^{}_{m} 
+ 
\mbox{H.c.} 
\right), \quad 
\psi^{}_{m} =  \! \int_{-D}^D \! d\epsilon \sqrt{\rho} 
\, c^{\phantom{\dagger}}_{\epsilon m}, \! 
\label{eq:H_V}
\\
&  \, E_d^{}= \  \epsilon_{d}^{} + \langle n_{dm}\rangle (N-1)U ,
\label{eq:Ed} 
\end{align}
Here, 
$d_m^{\dagger}$ is the creation operator for an electron 
with energy $\epsilon_{d}$ and orbital $m$ ($=1,2, \cdots, N$) 
in the impurity site, and 
$n_{dm}^{} =\, d_{m}^{\dagger} d_{m}^{}$. 
The operator $c_{\epsilon m}^{\dagger}$ 
creates a conduction electron  
that is normalized as 
$
\{ c^{\phantom{\dagger}}_{\epsilon m}, 
c^{\dagger}_{\epsilon'm'}
\} = \,\delta_{mm'}   
\delta(\epsilon-\epsilon')$.
The hybridization energy scale is given by 
 $\Delta = \pi \rho\, V_{}^2$  with $\rho=1/(2D)$.
We consider the parameter region where 
$\max(\Delta,\, |\epsilon_d|,\,U) \ll D$.

We use the imaginary-frequency Green's function 
that takes the form for $|\omega| \ll D$,
\begin{align}
G(i\omega) 
\,=\,  \frac{1}{i\omega - E_d^{} 
+ i \Delta \, \mathrm{sgn}\,\omega 
- \Sigma (i\omega)} \;,
\label{eq:G_dd}
\end{align}
where 
$\Sigma (i\omega)$ is the self-energy due to ${\cal H}_U$.
The ground-state average of the local charge 
can be deduced from 
the phase shift $\delta \equiv  \cot^{-1} (E_d^*/\Delta)$  
 with $E_d^* \equiv  E_{d}+\Sigma(0)$, 
using the Friedel sum rule  $\langle n_{dm} \rangle = \delta/\pi$.   
The derivative of  $\Sigma (i\omega)$ 
determines 
the renormalized parameters 
$\widetilde{\gamma} \equiv  
1 -  {\partial\Sigma(i\omega)}/{\partial (i\omega)}
|_{\omega=0}$, 
$z \equiv {1}/{\widetilde{\gamma}}$, 
$\,\widetilde{\epsilon}_{d} \equiv   z\,E_{d}^*$, 
and 
$\widetilde{\Delta}  \equiv z \Delta$. 
Furthermore, using the Ward identities,\cite{Yoshimori}  
the enhancement factor for spin 
and charge  susceptibilities, 
$\widetilde{\chi}_{s}^{} \equiv
\widetilde{\chi}_{mm}^{} - 
\widetilde{\chi}_{mm'}^{}$ and 
$\widetilde{\chi}_{c}^{}  \equiv 
\widetilde{\chi}_{mm}^{} + (N-1)\, \widetilde{\chi}_{mm'}^{}$, 
can be expressed in terms 
of the renormalization factor $\widetilde{\gamma}$ and 
the vertex function  
 $\Gamma_{mm';m'm}^{}(i \omega_1,i\omega_2;i\omega_3,i\omega_4)$ 
for $m\neq m'$,
\begin{align}
\widetilde{\chi}_{mm}^{} = \widetilde{\gamma} , 
\quad \ \ 
\widetilde{\chi}_{mm'}^{} =    -\,
\frac{\sin^2 \! \delta}{\pi\Delta}
\, \Gamma_{mm';m'm}^{}(0,0;0,0) .
\label{eq:ward_fermi}
\end{align}
The residual interaction between the quasi-particles 
is also given by  $\widetilde{U} 
\equiv z^2 \Gamma_{mm';m'm}^{}(0,0;0,0)$ for $m \neq m'$.

We showed in the previous work
that it is essential for $N>2$ to scale the Coulomb interaction   
by multiplying a factor $N-1$ 
in the particle-hole symmetric case.\cite{AoSakanoFujii}
Away from half-filling, it is more natural to 
include an additional factor $\sin^2 \! \delta$ such that
\begin{align} 
K^* \equiv g\,\sin^2 \! \delta \;, 
\qquad \qquad 
g \equiv \frac{(N-1)\,U}{\pi\Delta} \;,
\end{align} 
as the local density of states at the impurity site is given by   
 $- \frac{1}{\pi}\,\mathrm{Im}\,G(i0^+) = \sin^2 \! \delta/(\pi \Delta)$. 
Similarly, 
we introduce the renormalized version of the 
scaled coupling constant,
\begin{align}
\widetilde{K} \,\equiv\,  \widetilde{g}\,\sin^2 \! \delta  
\;,\qquad \quad 
\widetilde{g} \,\equiv\, 
\frac{ (N-1)\,\widetilde{U}}{\pi \widetilde{\Delta}} 
\;. 
\label{eq:g_scale}
\end{align}
The Wilson ratio $R$ and that for the charge sector 
can be expressed in terms of 
the renormalized coupling  $\widetilde{K}$,
\begin{align}
R  \equiv \frac{\widetilde{\chi}_{s}}{\widetilde{\gamma}}
 =  1+\frac{\widetilde{K}}{N-1}  
,
\qquad \quad 
\frac{\widetilde{\chi}_{c}}{\widetilde{\gamma}}
 =   1- \widetilde{K} .
\label{eq:Wilson}
\end{align}
Conversely, it can also be written as $\widetilde{K}=(N-1)(R-1)$.
By definition, $\widetilde{K}$
is bounded in the range  $0\leq \widetilde{K} \leq 1$, 
and  approaches $\widetilde{K} \to 1$  
when the charge fluctuation is suppressed $\widetilde{\chi}_{c}\to 0$.
At half-filling $\epsilon_d =-(N-1)U/2$ and $E_d=0$, 
it has been confirmed that 
$\widetilde{K}$ varies in a narrow region as $N$ increases, 
and converges rapidly to the RPA value  
that is asymptotically exact 
in the $N \to \infty$ limit.\cite{AoSakanoFujii}

Away from half-filling, the renormalized impurity level $E_d^*$ 
determines the charge distribution near the impurity, 
and should be calculated in an optimal way. 
We treat this problem by taking the unperturbed Green's function such that 
\begin{align}
G_0(i\omega) 
\,= \, \frac{1}{i\omega -E_d^* + i \Delta \,\mathrm{sgn}\, \omega} \;. 
\end{align}
In this set up, the energy shift due to $\Sigma(0)$ is included 
in the unperturbed part of the Hamiltonian, 
defined by ${\cal H}_0 =  {\cal H}_d + {\cal H}_c + {\cal H}_V 
+ \sum_m \Sigma(0) \,n_{dm}$,   
and it is subtracted in the perturbation Hamiltonian 
\begin{align}
{\overline{\cal H}_U}  =& \  {\cal H}_U - \sum_m \lambda \, n_{dm} \;. 
\end{align}
Here, the counter term is defined formally 
by $\lambda = \Sigma(0)$.\cite{Hewson} 
Then, the full Green's function defined in Eq.\ \eqref{eq:G_dd}
 can be rewritten in the form 
 \begin{align}
 G(i\omega) 
 \,=\, \frac{1}{i\omega -E_d^* + i \Delta
 \,\mathrm{sgn}\, \omega 
  - \overline{\Sigma}(i\omega)}   
\;. 
\end{align}
The redefined self-energy $\overline{\Sigma}(i\omega)$ represents the 
corrections due to ${\overline{\cal H}_U}$, 
and can be calculated perturbatively 
as a function of $E_d^*$ and $\lambda$.
These two parameters 
can be determined through the renormalization condition  
\begin{align}
\overline{\Sigma}(0) = 0  \;,
\label{eq:renormalization_condition}
\end{align}
and Eq.\ \eqref{eq:Ed} that can be rewritten,  
using the Friedel sum rule, as
\begin{align}
\frac{\xi_d}{\Delta}  
\,=\,  \frac{E_d^* -\lambda}{\Delta} 
+\,g\,  \tan^{-1} \!\left(\frac{E_d^*}{\Delta}\right) \;,
\label{eq:xi_d_away2}
\end{align}
where  $\xi_d \equiv \epsilon_d +(N-1)U/2$. 
At half-filling $\xi_{d}^{} =0$,
both of the two parameters vanish, $E_d^*=0$ and  $\lambda=0$.

\begin{figure}[t]
 \leavevmode
\begin{minipage}{1\linewidth}
\includegraphics[width=0.75\linewidth]{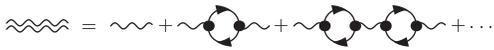}
\end{minipage}
\caption{The leading order diagrams in the $1/(N-1)$ expansion.
The wavy and solid lines indicate 
the Coulomb repulsion $U$ and 
unperturbed Green's function $G_0$, respectively.
The double wavy line represents the sum of the bubble diagrams, 
which corresponds to $\mathcal{U}_\mathrm{bub}(i\omega)$ 
given in  Eq.\  \eqref{eq:vertex_rings}. 
Each fermion loop gives a factor of order $(N-1)$ 
through a sum over the different orbitals 
[See Appendix \ref{sec:bubble_coefficient}]. 
}
 \label{fig:vertex_rings}
\end{figure}


\section{$1/(N-1)$ expansion}
\label{sec:1_(N-1)_expansion}

The perturbation expansion with respect to $\overline{\cal H}_U$   
can be classified according to the power of $1/(N-1)$,
by choosing the scaled interaction $g$ as an independent 
variable instead of bare $U$.
The counter term can also be expanded such that 
\begin{align}
\lambda = \sum_{k=0}^{\infty} \frac{\lambda_k}{(N-1)^k} \;,
\end{align}
and the coefficient $\lambda_k$ 
is determined by the requirement of the 
renormalization condition 
Eq.\ \eqref{eq:renormalization_condition}
is satisfied to each order in the $1/(N-1)$ expansion.

\subsection{Zeroth order}

\label{subsec:zeroth_order}

By construction, the  tadpole diagrams 
and the contributions arising from the bilinear part 
of $\mathcal{H}_U$ defined in  Eq.\ \eqref{eq:H_U},  i.e.,  
$-U\sum_{m\neq m'}\langle n_{dm'} \rangle\, n_{dm}$,  
cancel each other. 
Thus at zero order, the counter term vanishes $\lambda=0$, 
and $E_d^*$ is determined by Eq.\ \eqref{eq:xi_d_away2},  
which in this case gives the HF value,
\begin{align}
\frac{\xi_d}{\Delta}  
\,=\,  \frac{E_{d:\mathrm{HF}}^*}{\Delta} 
+\,g\,  \tan^{-1} \!\left(\frac{E_{d:\mathrm{HF}}^*}{\Delta}\right) \;.
\label{eq:HF}
\end{align}
Note that the corrections due to the RPA are still not 
included in the zeroth order.

\subsection{Leading order}
\label{subsec:leading_order}

\begin{figure}[t]
\begin{minipage}{1\linewidth}
\includegraphics[width=0.24\linewidth]{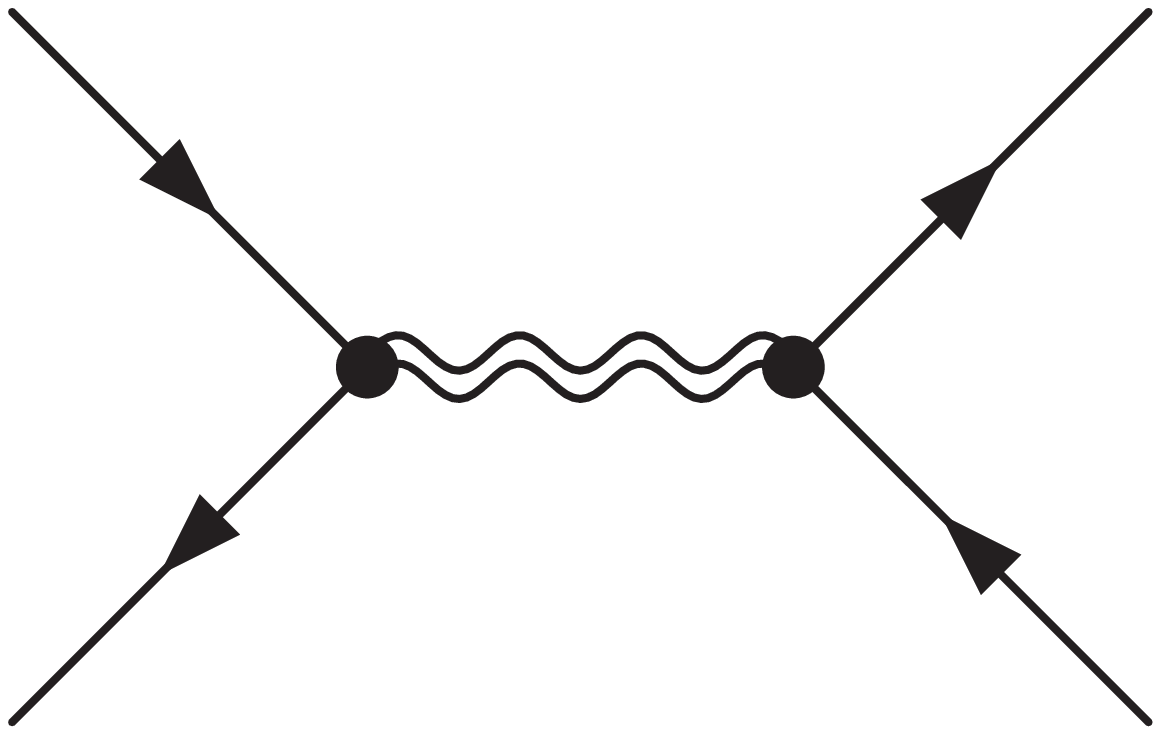}
\rule{0.06\linewidth}{0cm}
\includegraphics[width=0.35\linewidth]{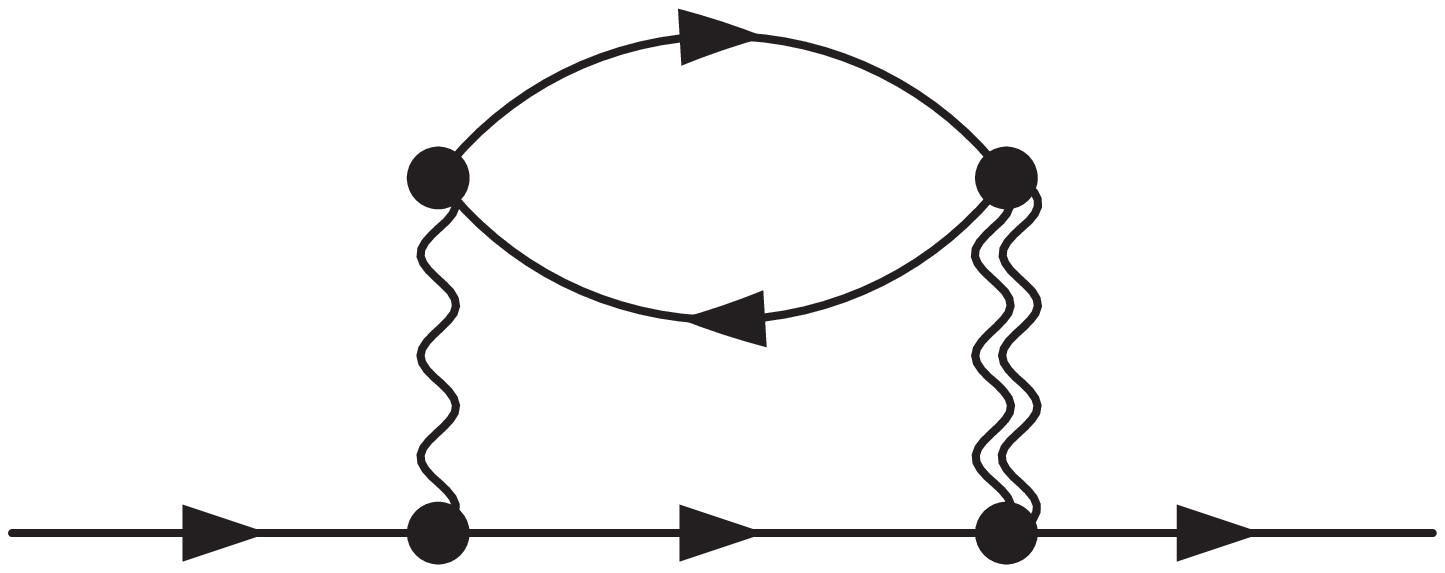}
\rule{0.02\linewidth}{0cm}
\raisebox{0.34cm}{
\includegraphics[width=0.22\linewidth]{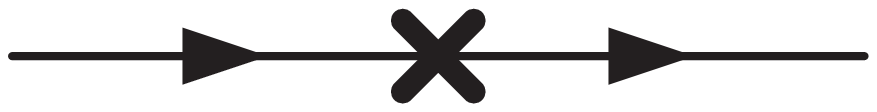}}
\end{minipage}
\caption{
The diagrams which provide the order $1/(N-1)$ contributions  
with higher order corrections. 
The cross denotes $-\lambda$ of the counter term.
}
 \label{fig:sg_rpa}
\end{figure}

The leading order corrections in the $1/(N-1)$ expansion 
arise form a series of the bubble diagrams of the 
RPA type indicated in Fig.\ \ref{fig:vertex_rings}, 
and can be expressed in the form  
\begin{align}
&
 \!\! 
\mathcal{U}_\mathrm{bub}^{}(i\omega) 
=   
\frac{\phi(i\omega)}{N-1} 
+ \frac{g \pi \Delta \, \Pi(i\omega)}{(N-1)^2} 
 +  O\!\left(\!\frac{1}{(N-1)^3}\!\right) , 
\label{eq:vertex_rings}
\\
&
\phi(i\omega) 
\equiv    
\frac{g \pi \Delta}{1+g \pi \Delta \chi_0^{}(i\omega)} 
, 
\label{eq:vertex_rpa2} 
\quad  
 \Pi(i\omega)
 \equiv  
 \chi_0^{}(i\omega)\,\phi(i\omega) .
\end{align}
The precise form of 
$\mathcal{U}_\mathrm{bub}^{}(i\omega)$ 
and $\chi_0^{}(i\omega)$ 
are provided in Appendix \ref{sec:bubble_coefficient}.  
The propagator $\mathcal{U}_\mathrm{bub}^{}(i\omega)$ 
 contains also the higher order contributions 
in the $1/(N-1)$ expansion. This is because the orbital indices 
for adjacent bubbles have to be different 
due to the Pauli exclusion, 
and thus the summations over 
internal $m$'s for the fermion loops are {\it not\/} independent. 
The order $1/(N-1)$ terms of $\Gamma_{mm';m'm}^{}(0,0;0,0)$ 
arises from the first diagram in Fig.~\ref{fig:sg_rpa},
which determines the leading order term of $\widetilde{K}$   
with the HF value of $K^*$,  as 
\begin{align}
\widetilde{K} =   \frac{K^*}{1+K^*} 
+O\!\left(\!\frac{1}{N-1}\!\right) .
\label{eq:g_inf}
\end{align}
Note that $\pi \Delta \chi_0^{}(0) =  
1/[1+(E_d^*/\Delta)^2]$ also equals $\sin^2 \! \delta$.
Equation \eqref{eq:g_inf} determines the Wilson ratio $R$,
defined in Eq.\ \eqref{eq:Wilson}, to order $1/(N-1)$.

The leading order self-energy due to ${\overline{\cal H}_U}$ 
arises from the second and third diagrams in Fig.\ \ref{fig:sg_rpa}.
These two diagrams represent the contributions of
the RPA fluctuation and counter term, 
and can be expressed in the form  
\begin{align}
&\overline{\Sigma}(i\omega) \,= \,  
\Sigma^{(1/N')}(i\omega) - \lambda  
\ +\, O\!\left(\!\frac{1}{N'^2}\!\right) ,
\\
&\Sigma^{(1/N')}(i\omega) 
\,= \,  
 \frac{g\,\pi \Delta}{N-1} \, 
\int \frac{d\omega'}{2\pi} 
 \,G_0^{}(i\omega-i\omega')\,
\Pi(i\omega') , 
\label{eq:sg_to_order2_w0}
\end{align}
where $N' =N-1$. 
Therefore,  
from the renormalization condition defined 
in Eq.\ \eqref{eq:renormalization_condition} 
the counter term can be deduced  to the first order in $1/(N-1)$, 
\begin{align}
\lambda   \,=\,  \Sigma^{(1/N')}(0) + O\!\left(\frac{1}{N'^2}\right) \;.
\label{eq:lambda_1}
\end{align}
In a similar way the coefficient $\lambda_k$ of 
the counter term can be determined order by order. 
We can also calculate the leading order term of $\widetilde{\gamma}$ 
from $\Sigma^{(1/N')}(i\omega)$, 
\begin{align}
& 
\!\!\!
\widetilde{\gamma}
= 1+ \frac{K^*}{N-1} 
\left[\frac{K^*}{1+K^*} + \mathcal{I}_{S}^{}\right] 
\! 
+ \widetilde{\gamma}^{(\frac{1}{N'^2})}
\! + \! 
 O\!\left(\!\frac{1}{N'^3}\!\right) \!, \!
\\
&
\! 
\mathcal{I}_{S}^{}
 =
  \left[1+\! \left(\!\frac{E_d^*}{\Delta}\!\right)^2\right]
\pi \Delta \! 
\int\! \frac{d\omega}{2\pi} 
 \left\{G_0^{}(i\omega)\right\}^2
\Pi(i\omega) .
\label{eq:sg_to_order2}
\end{align}
Here, $\widetilde{\gamma}^{(\frac{1}{N'^2})}$  denotes  
the next-leading order contributions, 
which will be taken into account later.

\subsection{Next leading order}
\label{subsec:next_leading_order}

\begin{figure}[t]
 \leavevmode
\begin{minipage}{1\linewidth}
\includegraphics[width=0.27\linewidth]{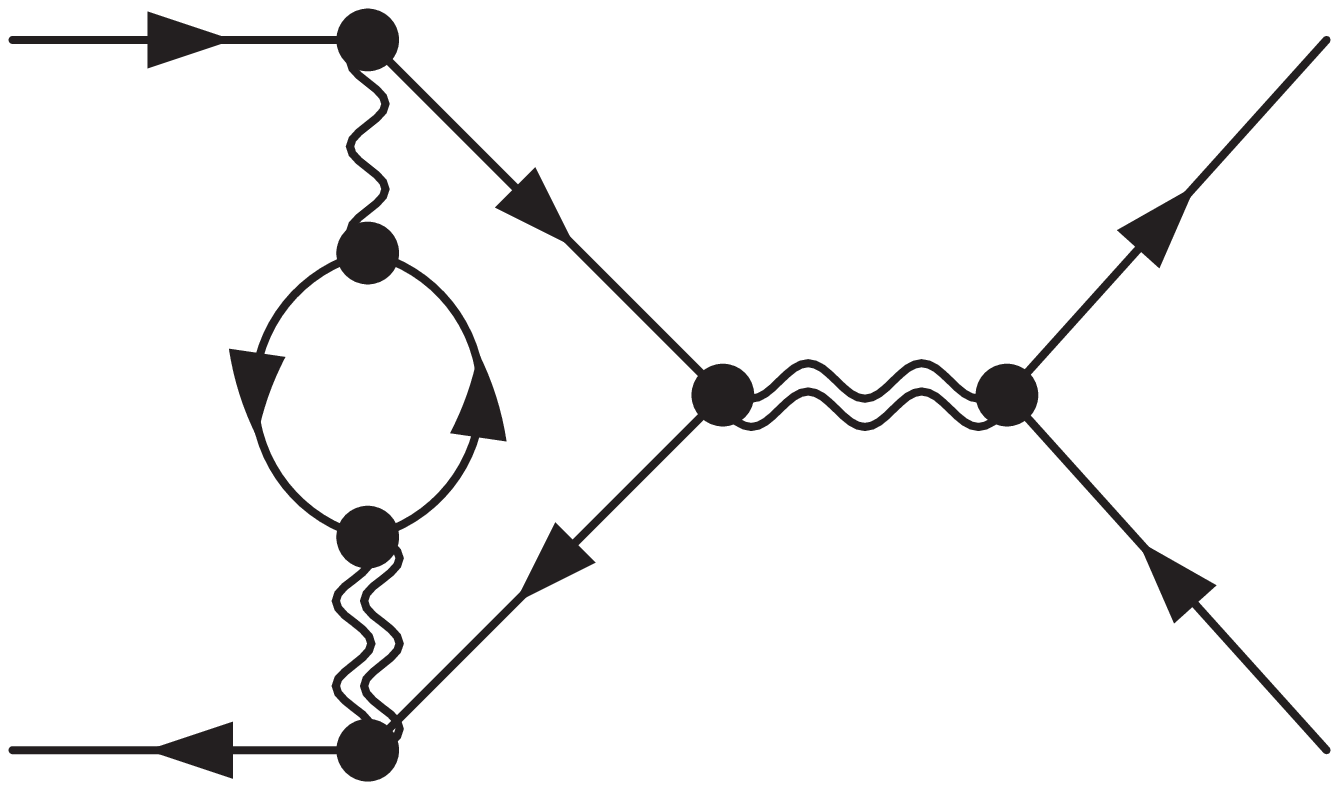}
 \rule{0.1\linewidth}{0cm}
\includegraphics[width=0.27\linewidth]{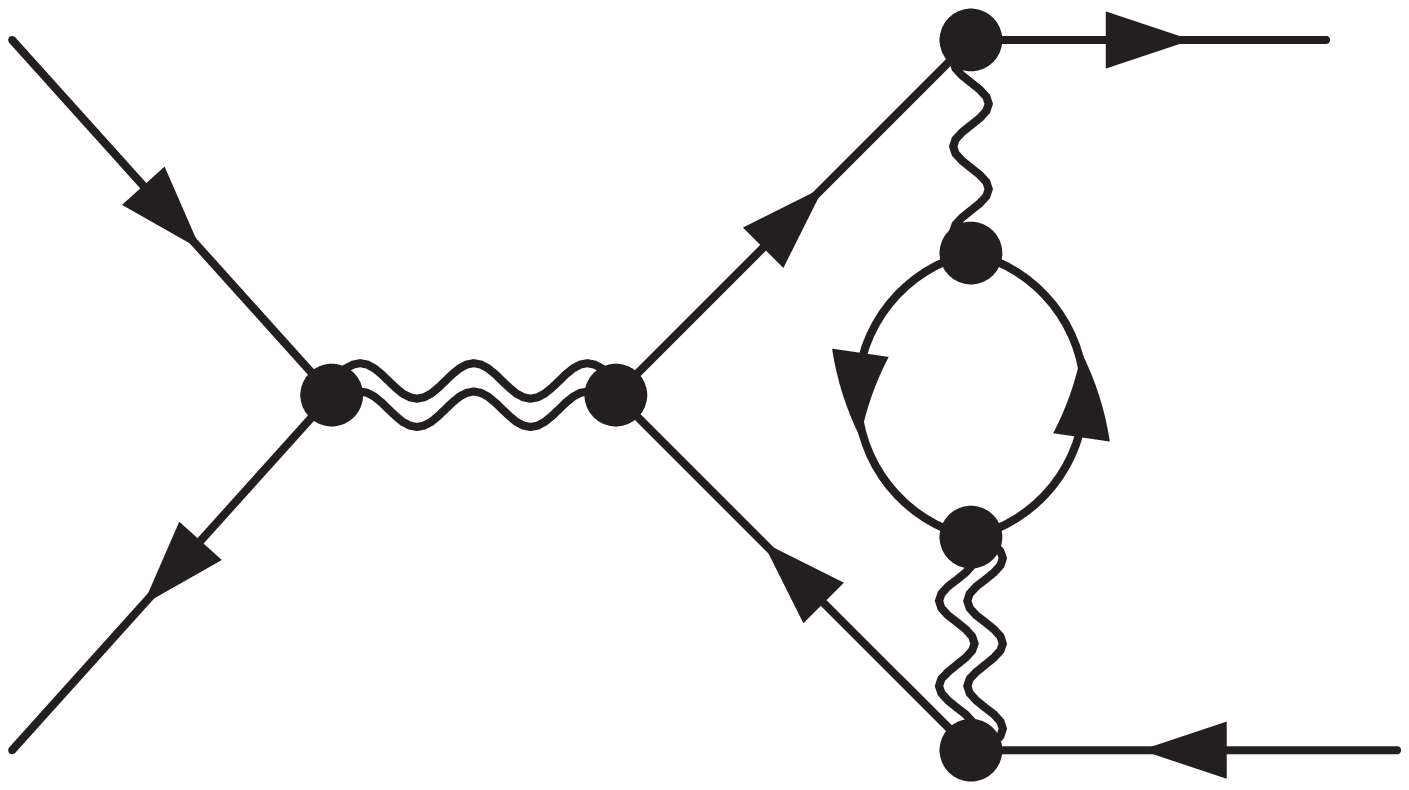}
\end{minipage}
 \rule{0cm}{0.25cm}
 \begin{minipage}{1\linewidth}
 \includegraphics[width=0.27\linewidth]{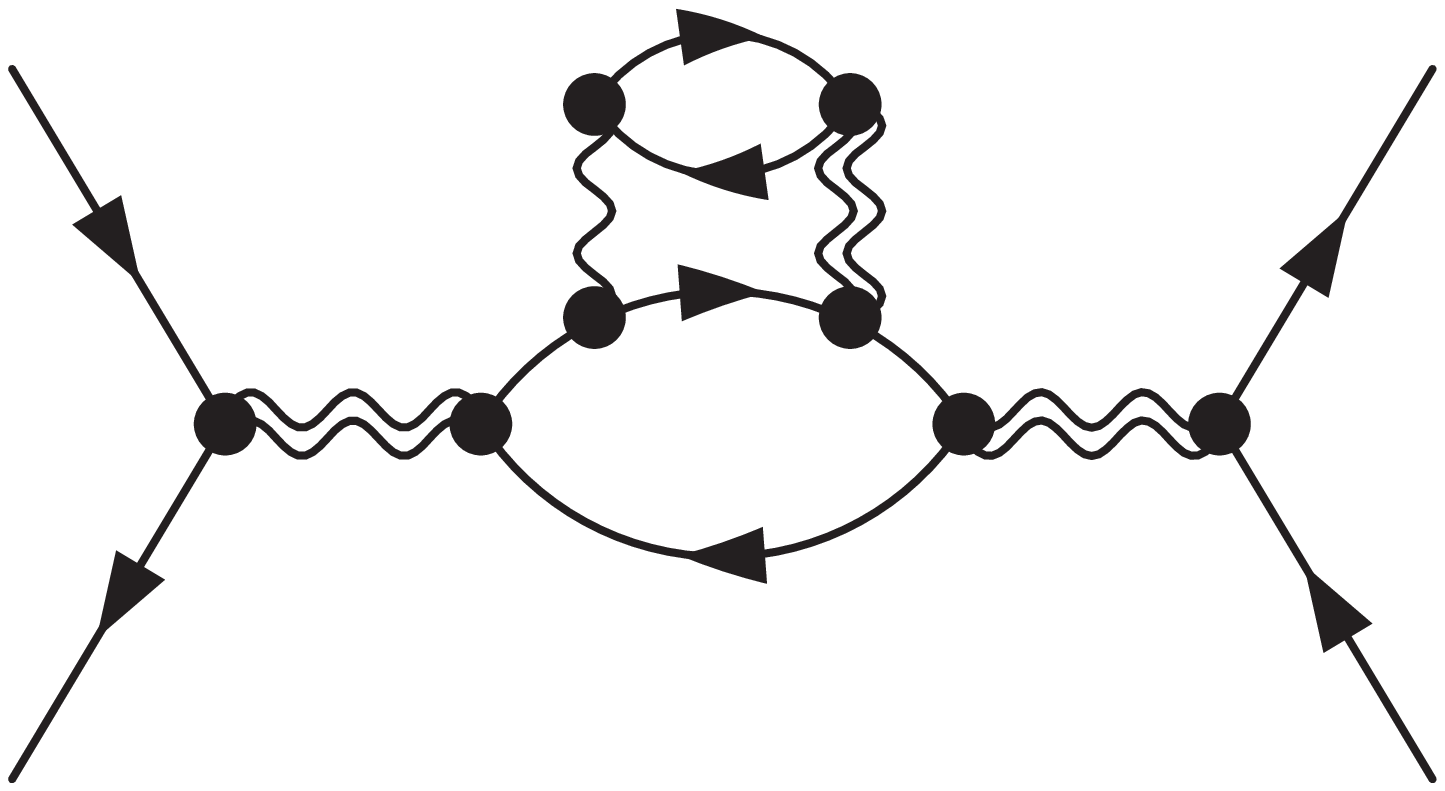}
 \rule{0.04\linewidth}{0cm}
 \includegraphics[width=0.27\linewidth]{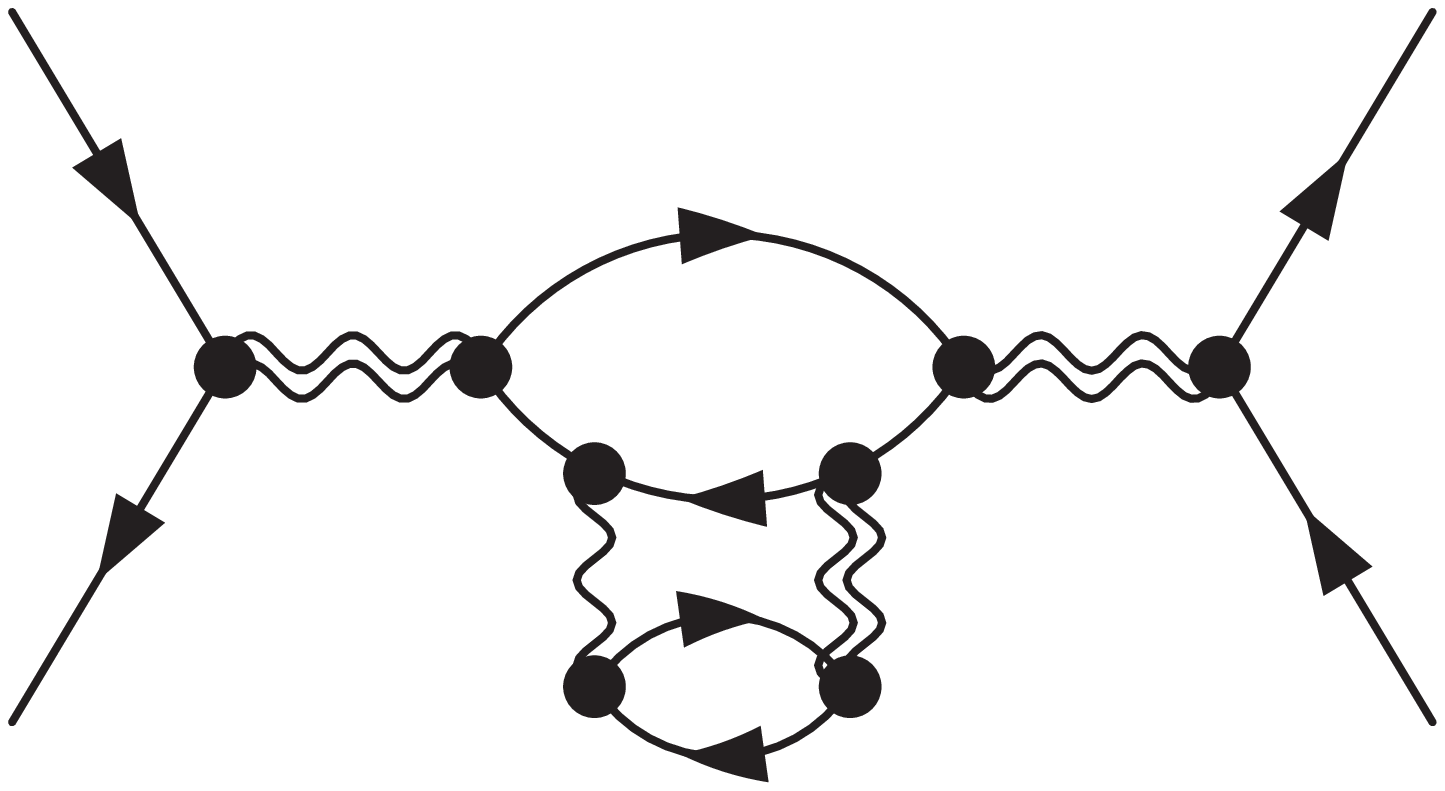}
 \rule{0.04\linewidth}{0cm}
 \includegraphics[width=0.27\linewidth]{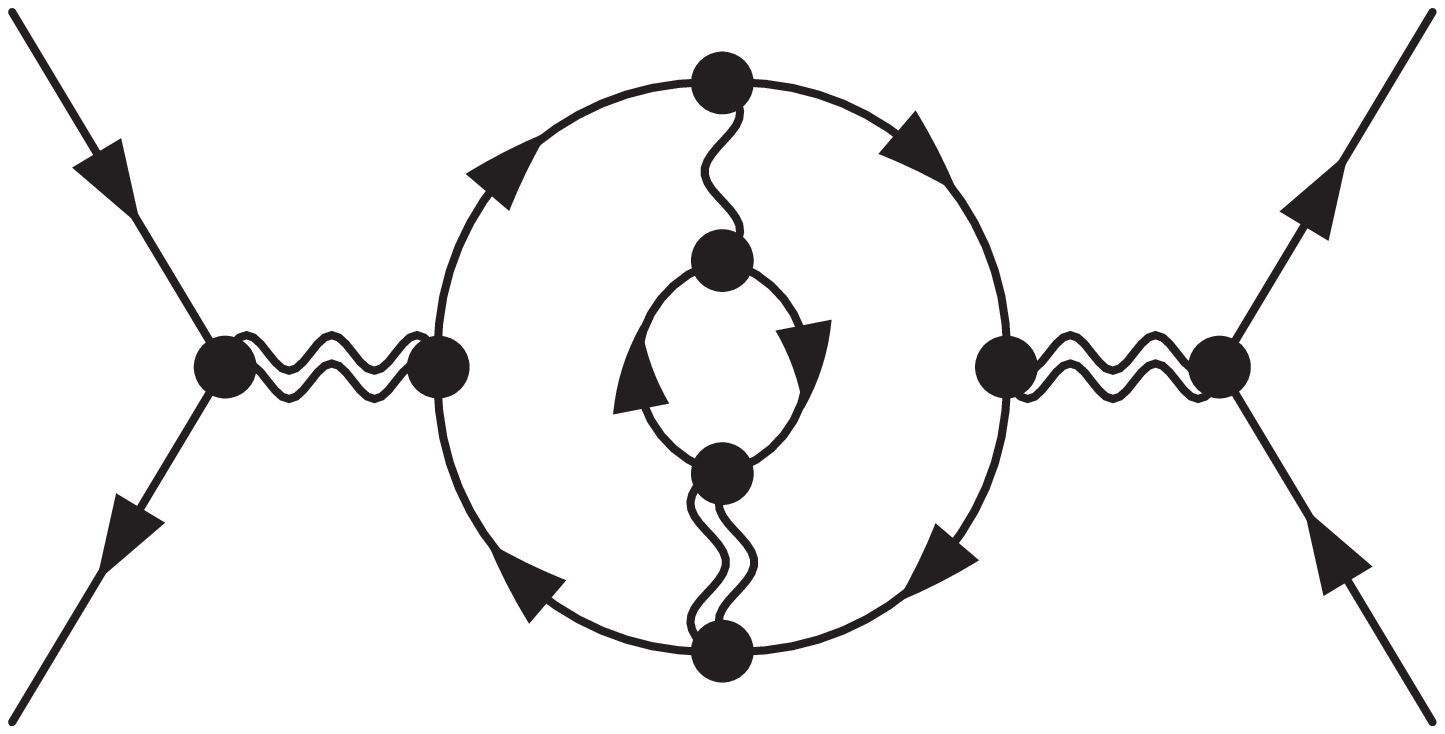}
 \end{minipage}
 \caption{
The order $1/(N-1)^2$ diagrams for the vertex function 
$\Gamma_{mm';m'm}^{}(0,0;0,0)$ for $m \neq m'$. 
}
 \label{fig:vertex_rpa}
\end{figure}

\begin{figure}[t]
 \leavevmode
\begin{minipage}{1\linewidth}
\includegraphics[width=0.23\linewidth]{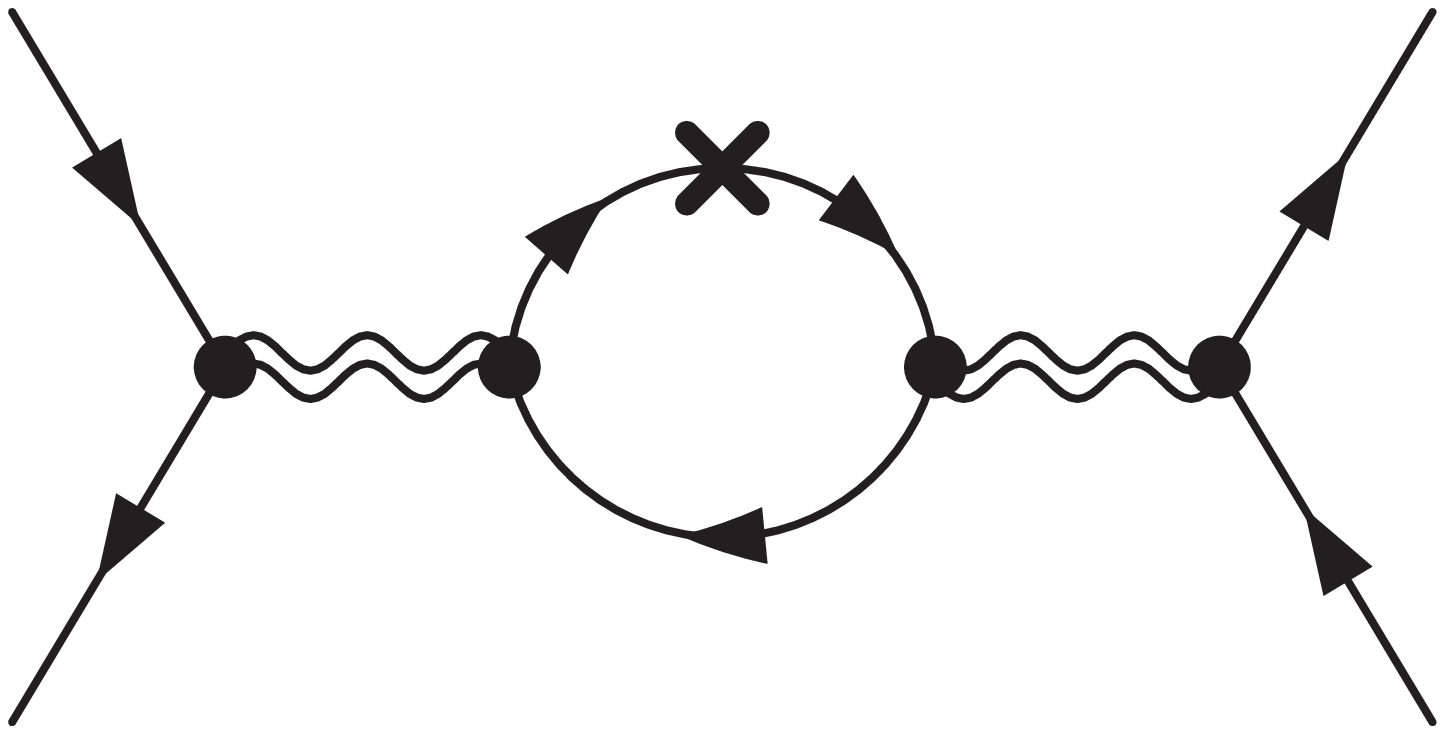}
 \rule{0.02\linewidth}{0cm}
\includegraphics[width=0.23\linewidth]{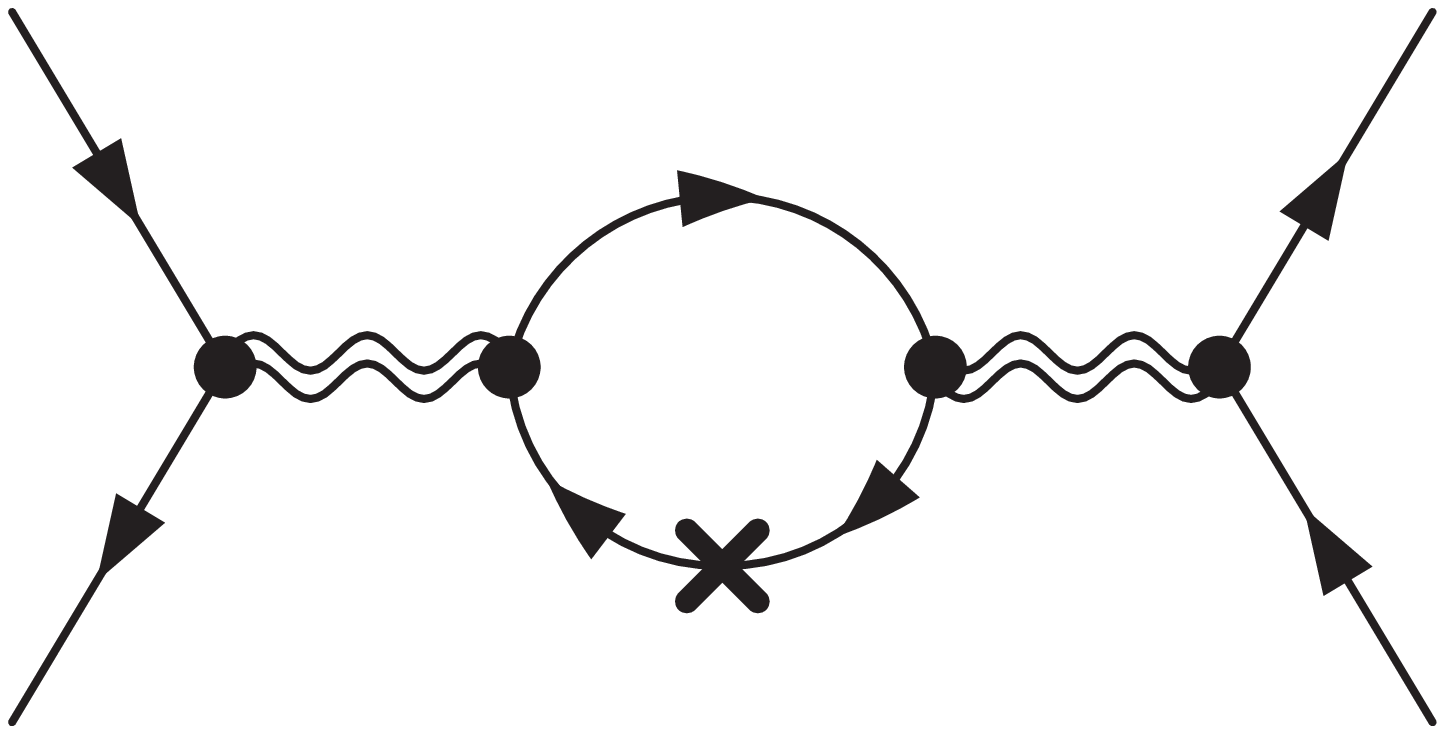}
 \rule{0.01\linewidth}{0cm}
\raisebox{0.1cm}{
\includegraphics[width=0.185\linewidth]{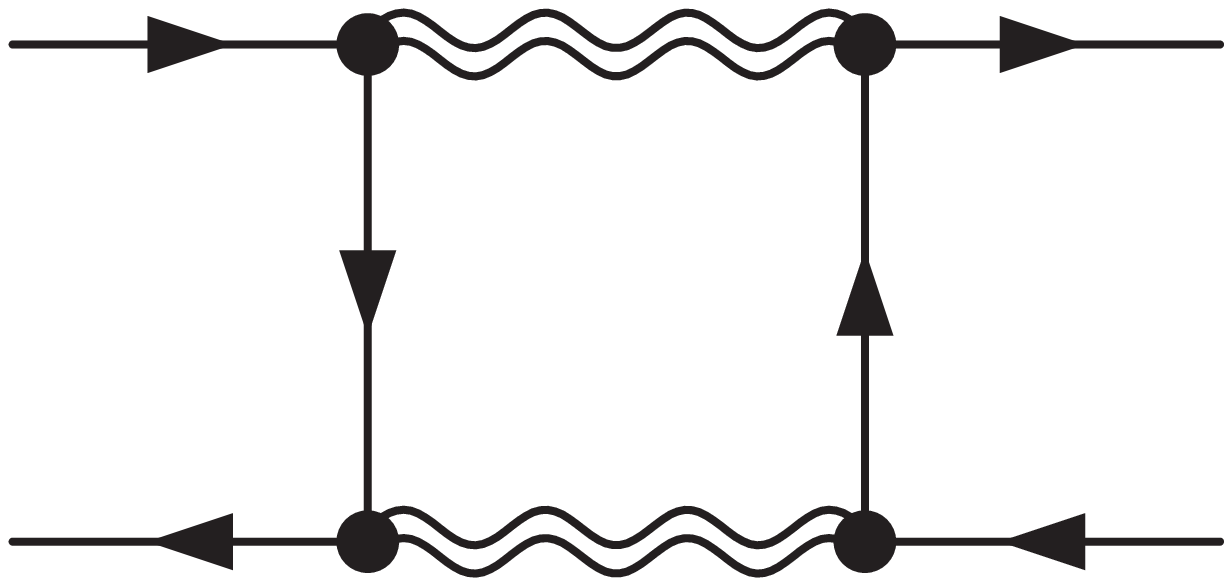}
 \rule{0.02\linewidth}{0cm}
\includegraphics[width=0.185\linewidth]{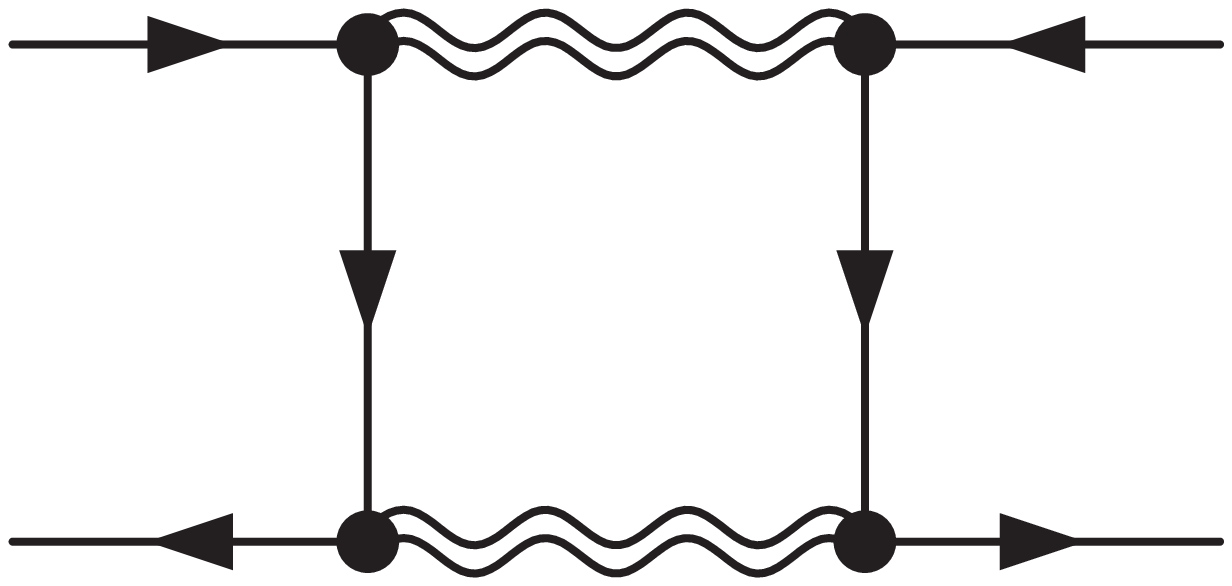}
}
\end{minipage}
 \rule{0cm}{0.28cm}
 \begin{minipage}{1\linewidth}
 \includegraphics[width=0.23\linewidth]{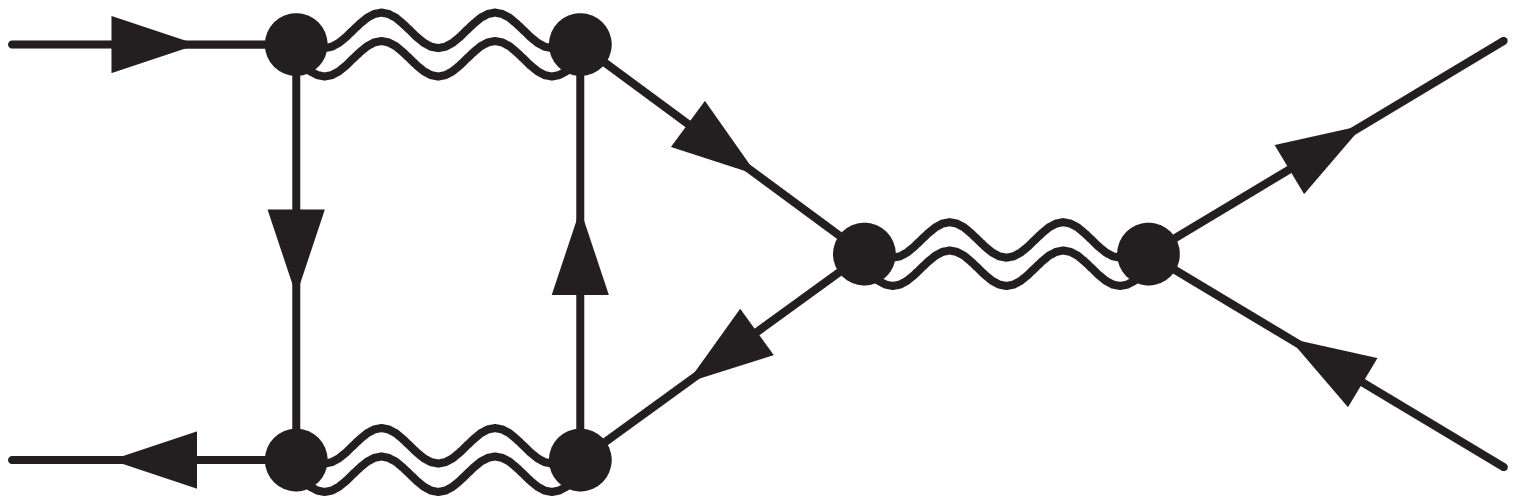}
 \rule{0.004\linewidth}{0cm}
 \includegraphics[width=0.23\linewidth]{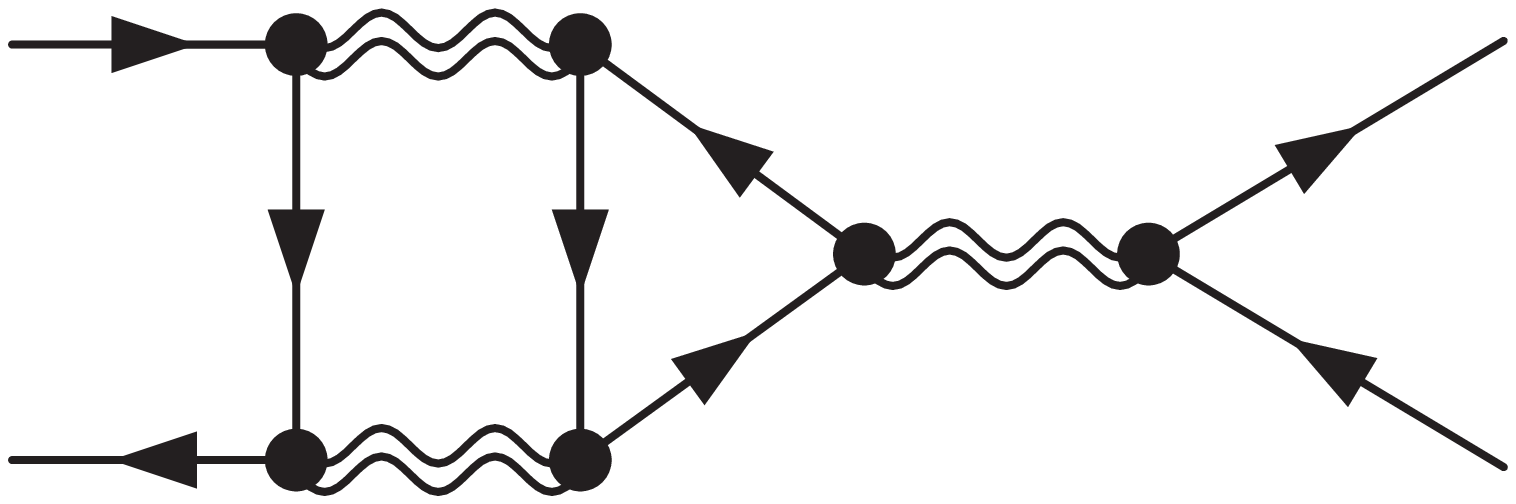}
 \rule{0.001\linewidth}{0cm}
 \includegraphics[width=0.23\linewidth]{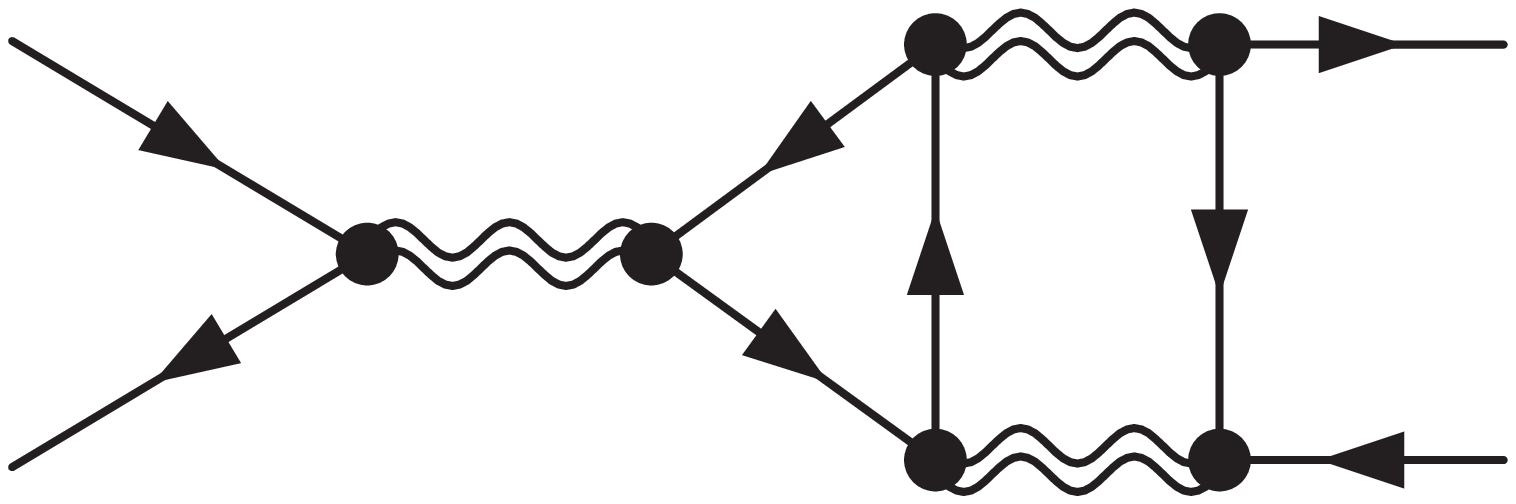}
 \rule{0.004\linewidth}{0cm}
 \includegraphics[width=0.23\linewidth]{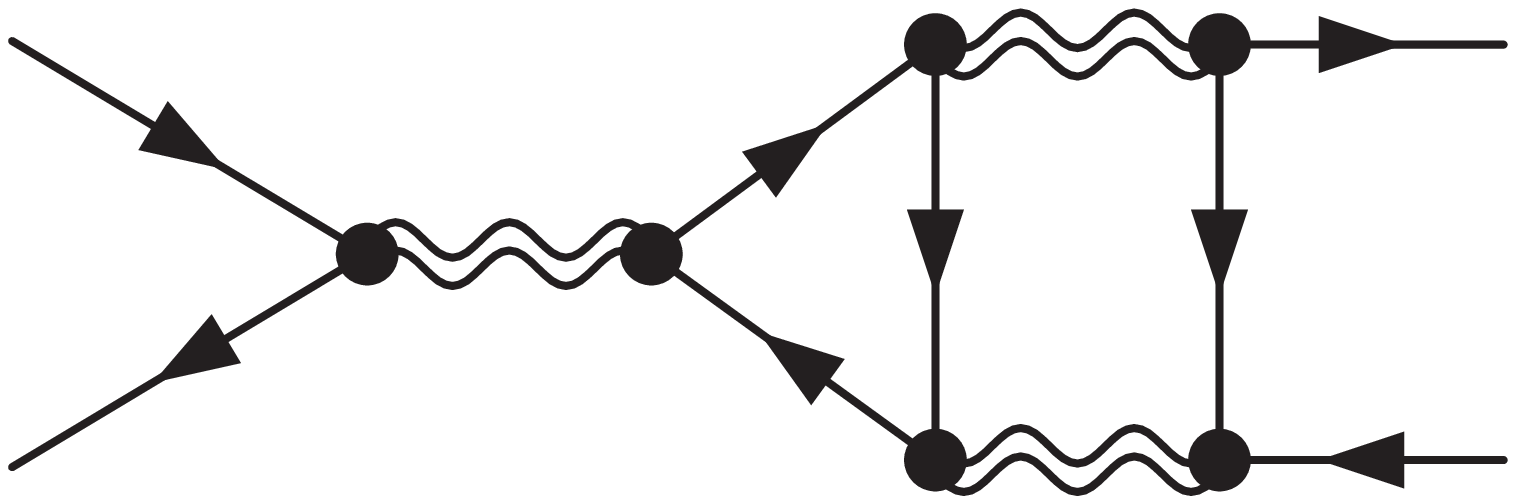}
 \end{minipage}
\rule{0cm}{0.1cm}
\begin{minipage}{1\linewidth}
\includegraphics[width=0.25\linewidth]{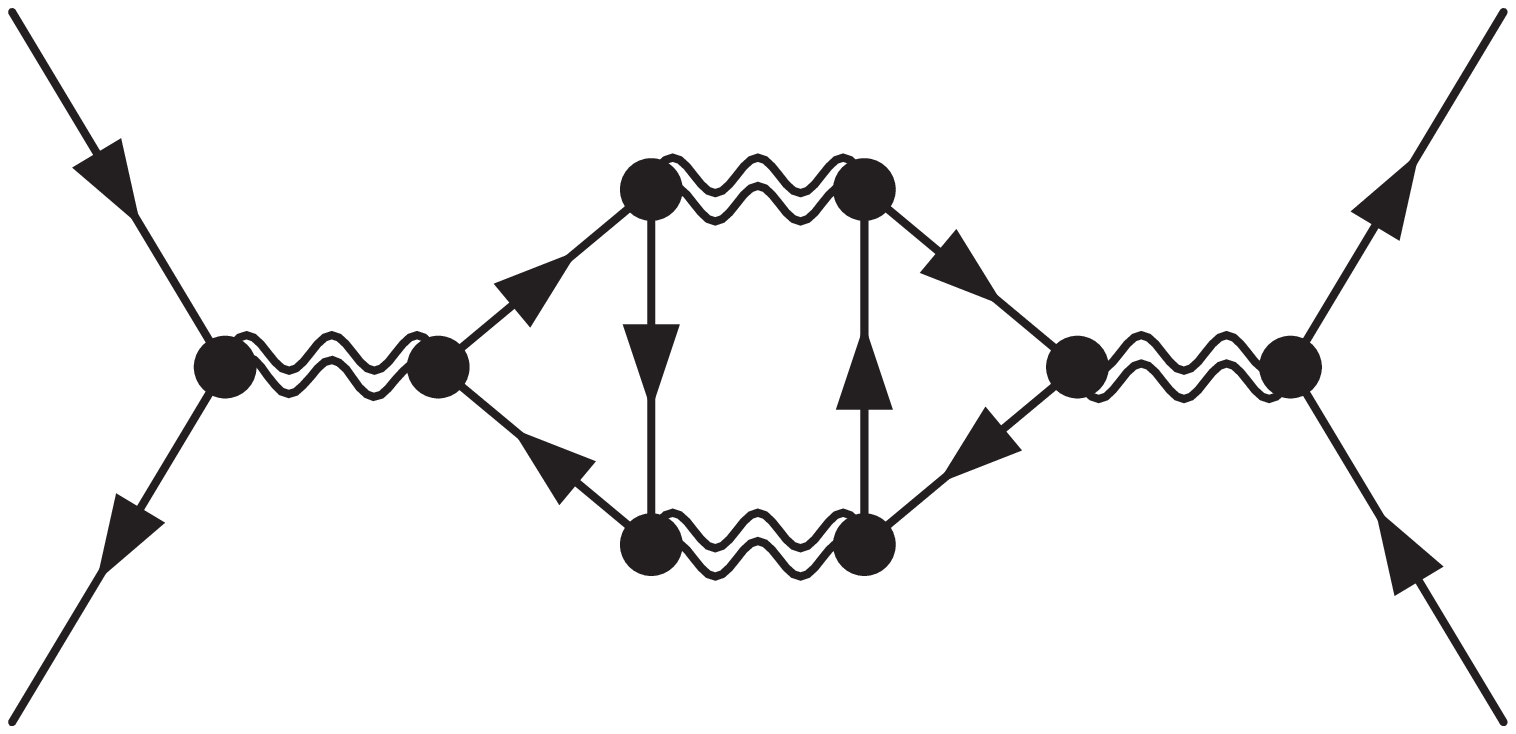}
 \rule{0.13\linewidth}{0cm}
\includegraphics[width=0.25\linewidth]{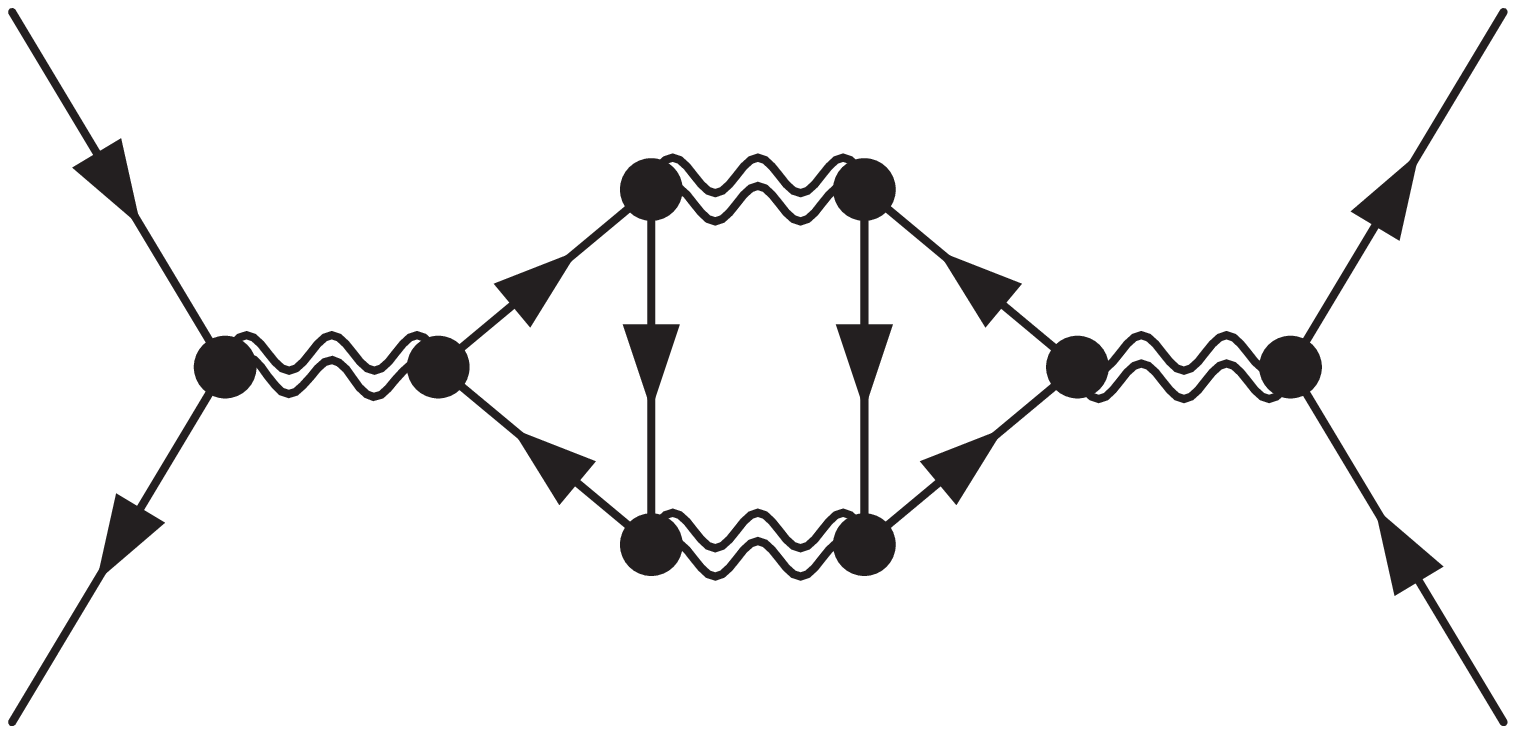}
\end{minipage}
 \caption{
The order $1/(N-1)^2$ vertex diagrams for $m \neq m'$   
contribute away from half-filling in addition to the ones 
given in Fig.\ \ref{fig:vertex_rpa}. 
The contribution of the particle-particle pairs
and that of particle-hole pairs cancel each other at half-filling.
}
 \label{fig:vertex_rpa_asmX}
\end{figure}

Fluctuations beyond the RPA 
enter  through the next-leading and 
higher order terms in the $1/(N-1)$ expansion. 
In the particle-hole {\it asymmetric\/} case, 
the order $1/(N-1)^2$ contributions to 
$\Gamma_{mm';m'm}^{}(0,0;0,0)$ arise from the diagrams  
shown in Figs.\ 
\ref{fig:vertex_rpa} and \ref{fig:vertex_rpa_asmX}, 
and also from the order $1/(N-1)^2$ order component of 
$\mathcal{U}_\mathrm{bub}^{}(i\omega)$ shown 
in the first diagram of Fig.\ \ref{fig:sg_rpa},
i.e.,  the second term  
in the right-hand side of Eq.\ \eqref{eq:vertex_rings}. 
Note that the contributions of the diagrams 
shown in Fig.\ \ref{fig:vertex_rpa_asmX}  
vanish  at half-filling.  
This is because the particle-particle 
and particle-hole pairs give opposite contributions, 
and cancel each other in the particle-hole symmetric case.

Summing up the contributions from all these diagrams, 
we calculate the Wilson ratio $R$, 
defined in Eq.\ \eqref{eq:Wilson}, to order $1/(N-1)^2$. 
The result of the renormalized coupling 
$\widetilde{K}$ can be expressed in the following form, 
which is exact up to terms of order $1/(N-1)$, 
\begin{align}
&
\!\!
\widetilde{K} =   
\frac{K^*}{1+K^*} 
\frac{1+ 
\frac{K^*}{N-1} \! 
\left[1+\left(2 - \frac{K^*}{1+K^*}\right)
\mathcal{I}_{S}^{}
- \frac{1}{1+K^*}\, \mathcal{I}_{A}^{}
\right]
}{1 +  \frac{K^*}{N-1} 
\left[\frac{K^*}{1+K^*} + \mathcal{I}_{S}^{}\right]
}  ,
\label{eq:vertex_rpa_correction}
\\
&
\!\!
\mathcal{I}_{A}^{}
 = 
 \left[1+\!\left(\!\frac{E_d^*}{\Delta}\!\right)^2\right]\!
\frac{\pi\Delta}{2} \! 
\int\!  \frac{d\omega}{2\pi} \! 
 \left[\frac{G_0^{}(i\omega)+G_0^{}(-i\omega)}
{1+g\pi \Delta \chi_0(i\omega)}\right]^2 \!\! . \!
\end{align}
In this formula 
the contribution described by $\mathcal{I}_{A}^{}$, emerging in the numerator 
in the right-hand side of Eq.\ \eqref{eq:vertex_rpa_correction}, 
becomes finite in the particle-hole {\it asymmetric\/} case,
and it represents the contributions from 
the eight diagrams in Fig.\ \ref{fig:vertex_rpa_asmX}, 
excluding the first two caused by the counter term.
The prefactor $K^*/(1+K^*)$ in the right-hand side of 
Eq.\ \eqref{eq:vertex_rpa_correction} is 
required to be calculated to order $1/(N-1)$,   
by including the energy shift due to Eq.\ \eqref{eq:sg_to_order2_w0}.
Alternatively, through Eq.\ \eqref{eq:xi_d_away2} 
with Eq.\ \eqref{eq:lambda_1}, 
the bare parameter $\xi_d$ can be determined 
as a function of $E_d^*$ to order $1/(N-1)$.

\begin{figure}[t]
 \leavevmode
\begin{minipage}{1\linewidth}
\includegraphics[width=0.24\linewidth]{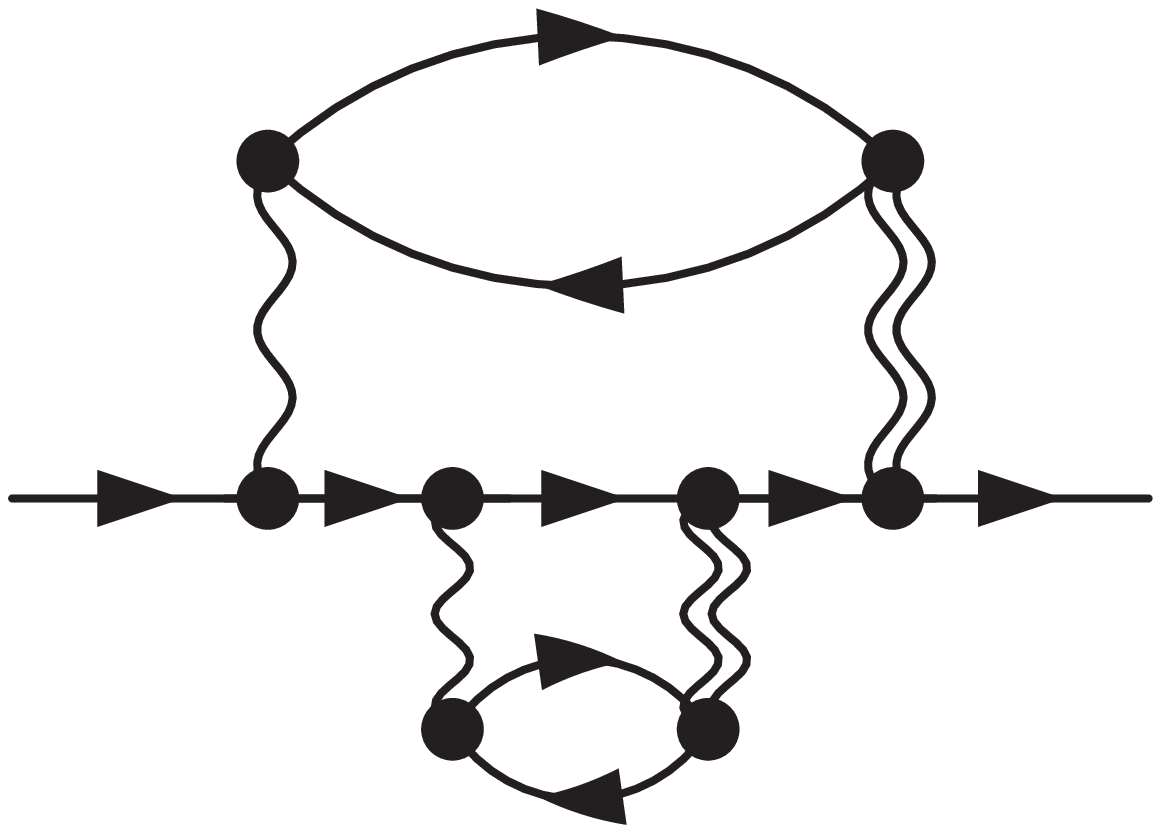}
 \rule{0.1\linewidth}{0cm}
\raisebox{-0.3cm}{
\includegraphics[width=0.26\linewidth]{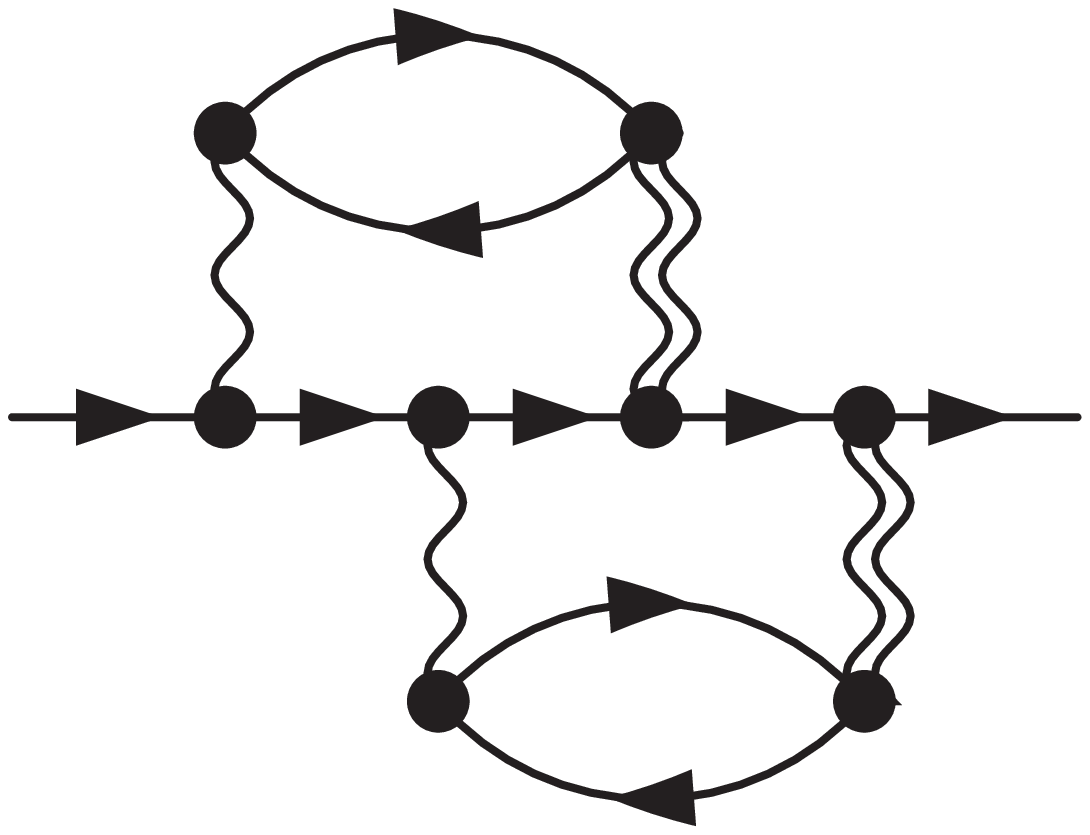}
}
\end{minipage}
 \begin{minipage}{1\linewidth}
 \includegraphics[width=0.27\linewidth]{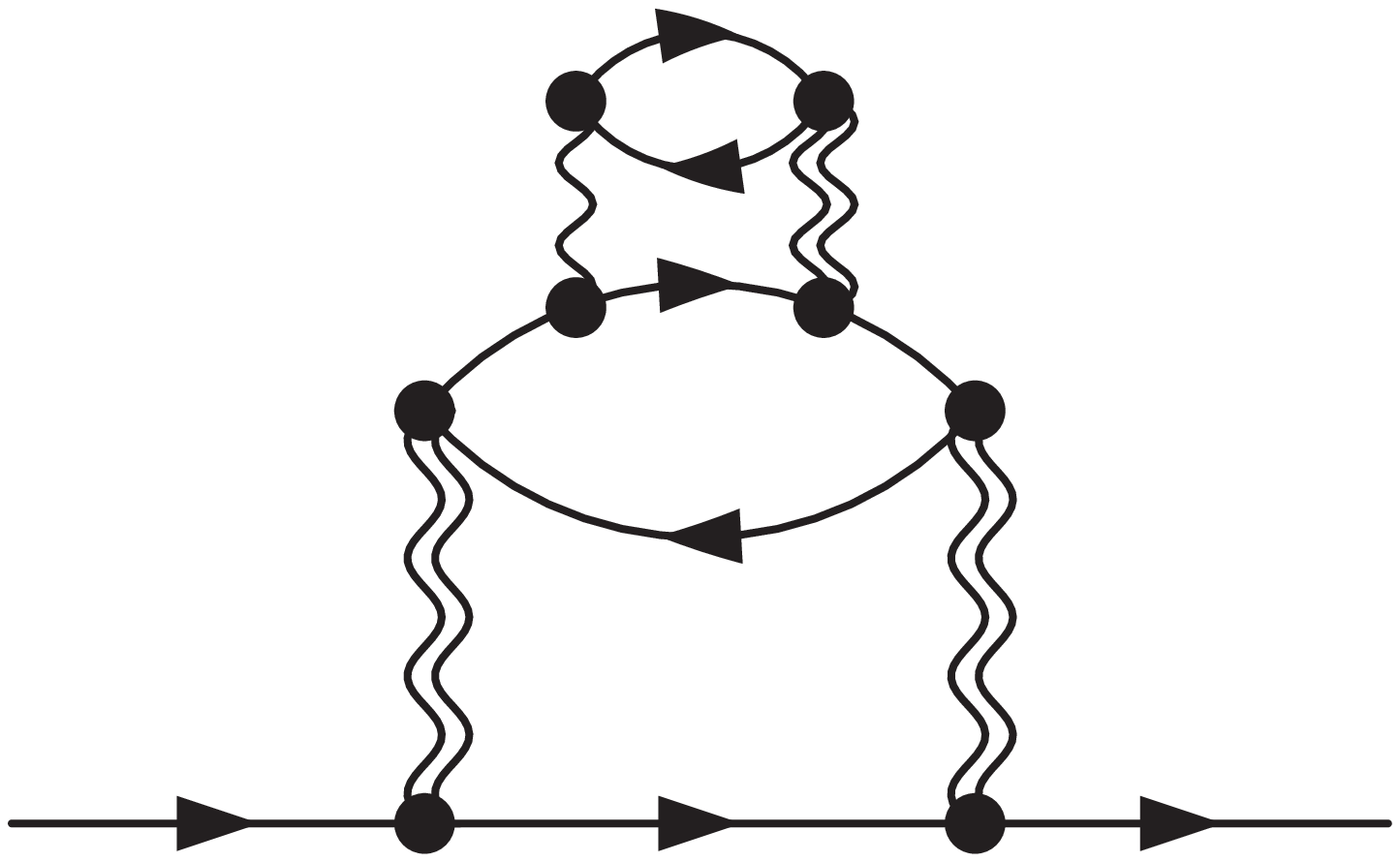}
 \rule{0.04\linewidth}{0cm}
 \includegraphics[width=0.27\linewidth]{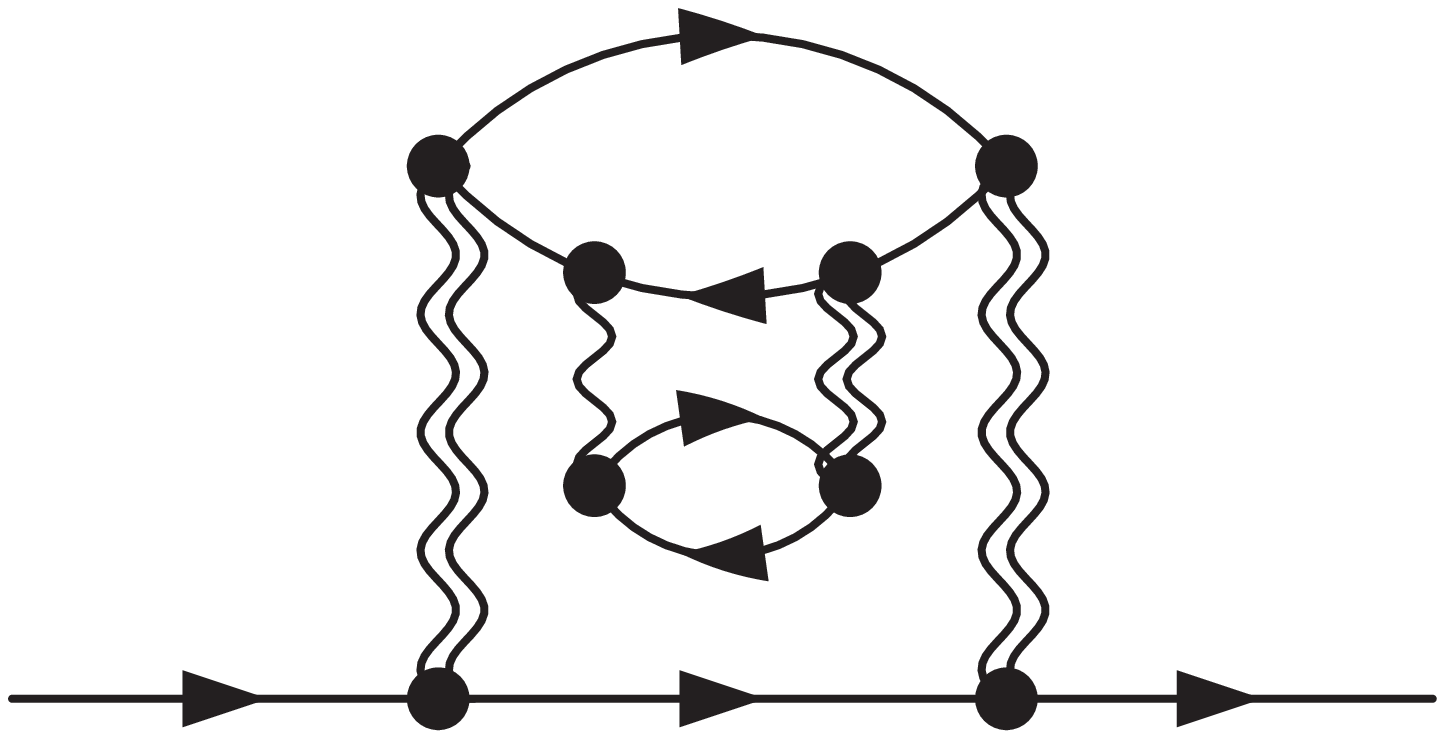}
 \rule{0.04\linewidth}{0cm}
 \includegraphics[width=0.27\linewidth]{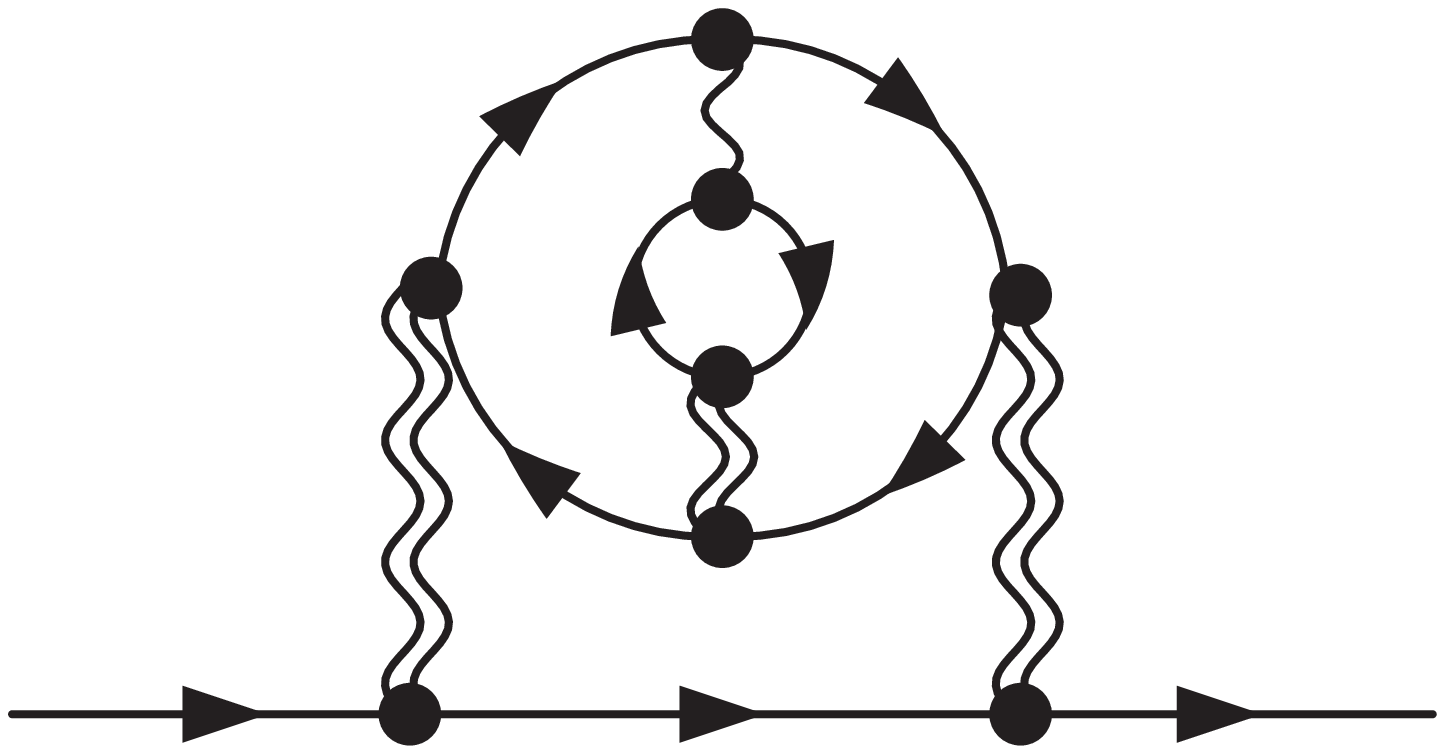}
 \end{minipage}
 \caption{
The order $1/(N-1)^2$ self-energy diagrams contribute to  
$z$ ($=1/\widetilde{\gamma}$).
}
\label{fig:sg_rpa_n2}
\end{figure}

\begin{figure}[t]
 \leavevmode
\begin{minipage}{1\linewidth}
\includegraphics[width=0.26\linewidth]{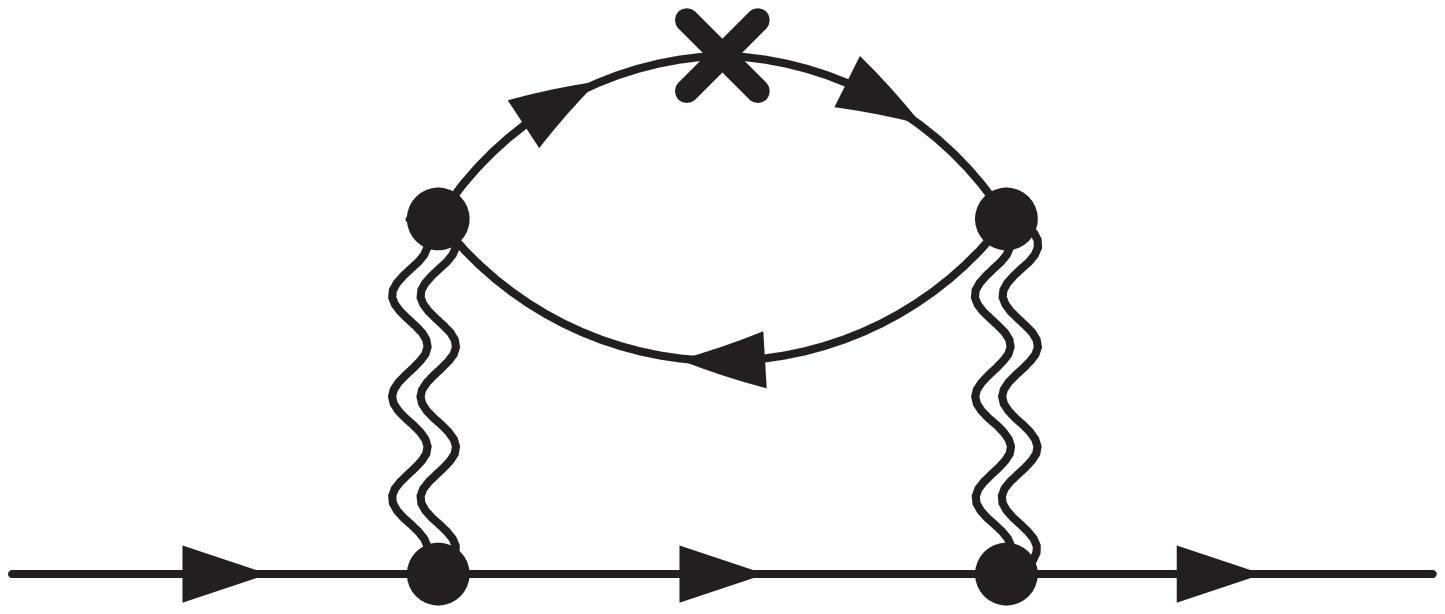}
 \rule{0.03\linewidth}{0cm}
\includegraphics[width=0.26\linewidth]{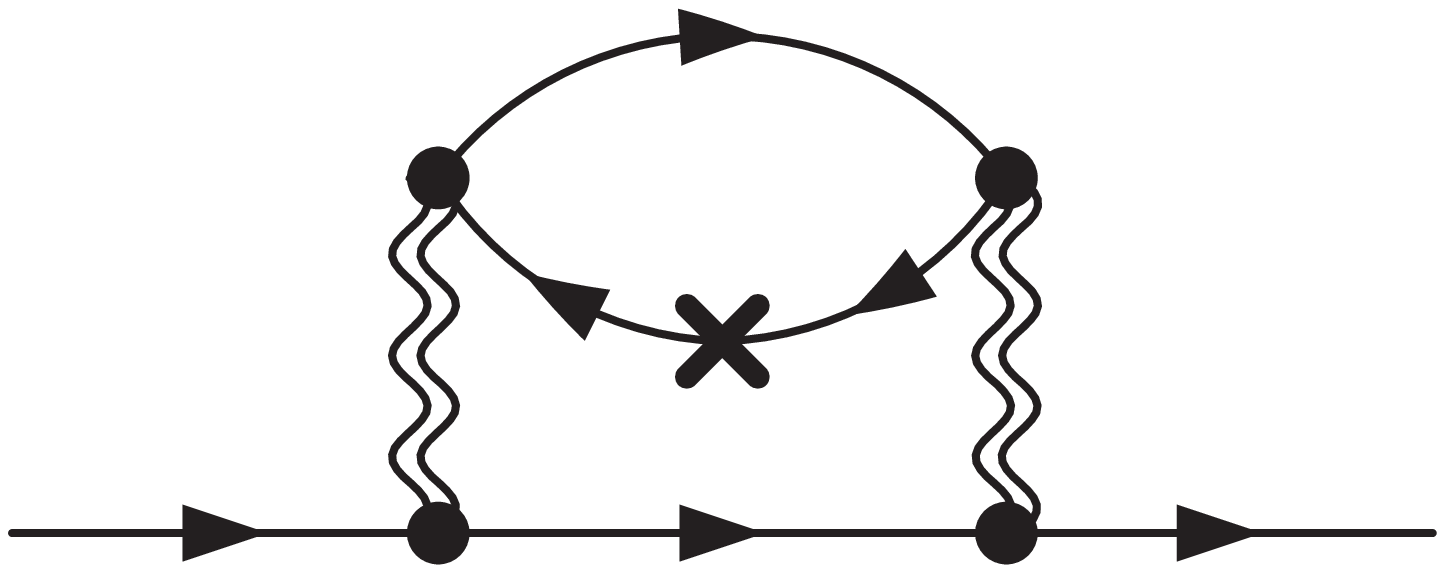}
 \rule{0.03\linewidth}{0cm}
\includegraphics[width=0.26\linewidth]{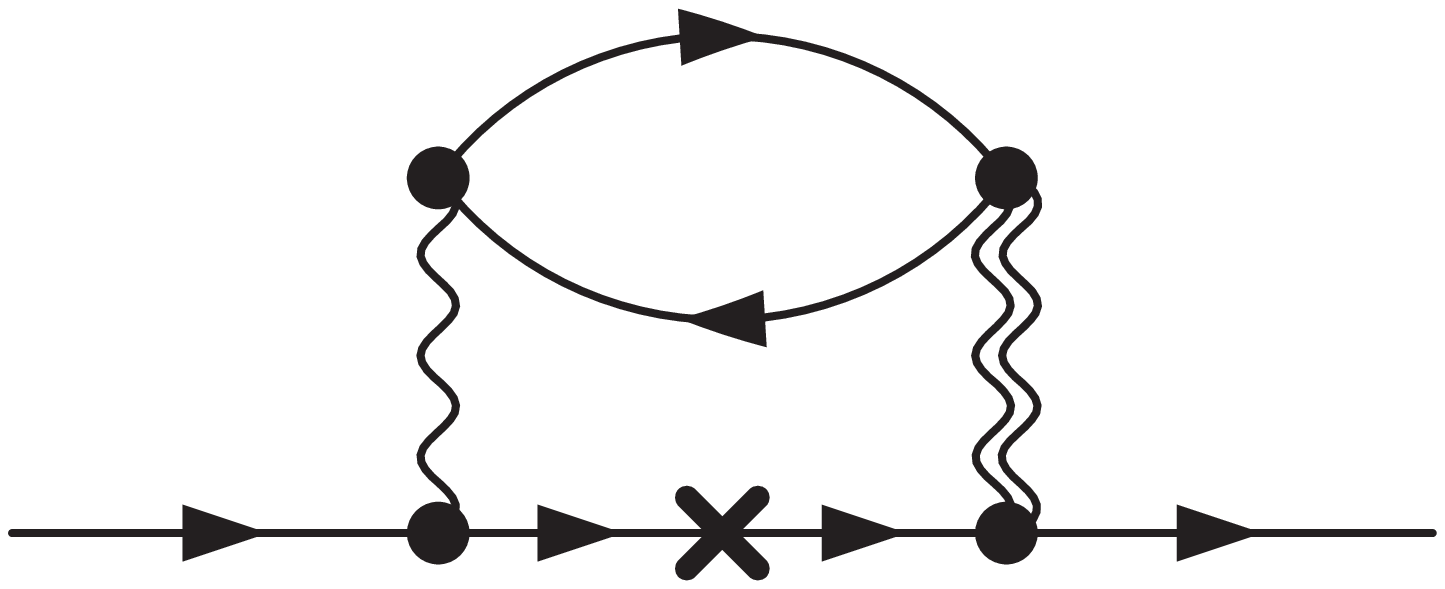}
\end{minipage}
 \rule{0.0cm}{0.18cm}
\begin{minipage}{1\linewidth}
\includegraphics[width=0.22\linewidth]{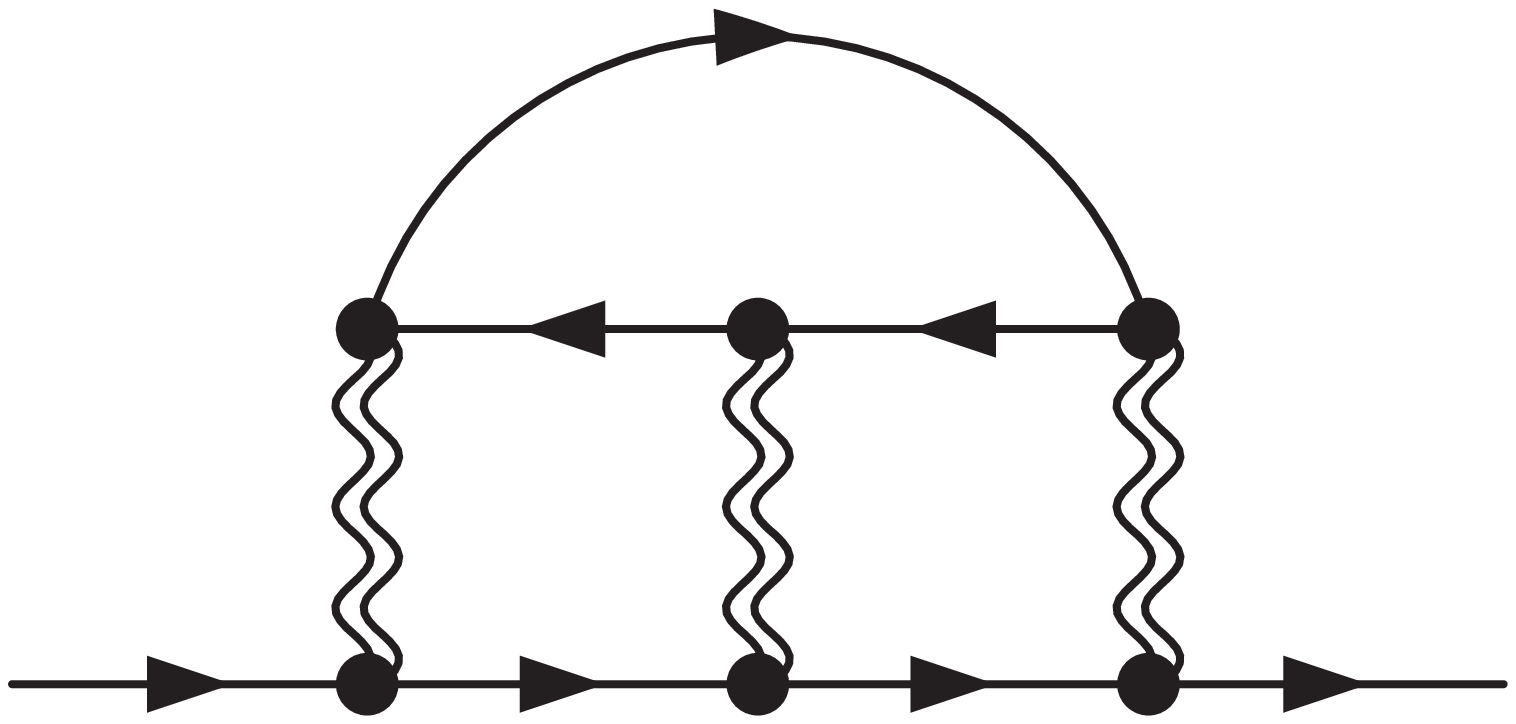}
 \rule{0.004\linewidth}{0cm}
\includegraphics[width=0.22\linewidth]{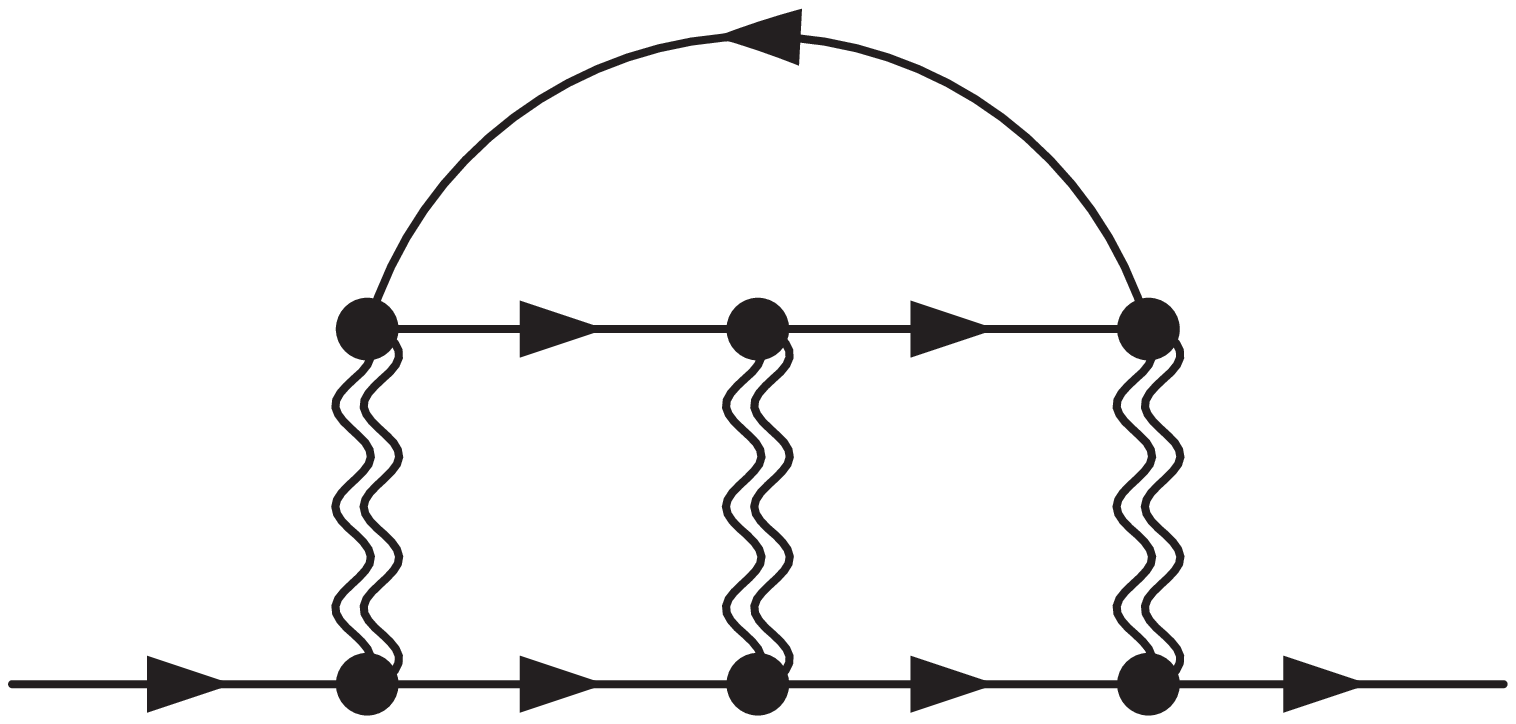}
 \rule{0.004\linewidth}{0cm}
 \includegraphics[width=0.24\linewidth]{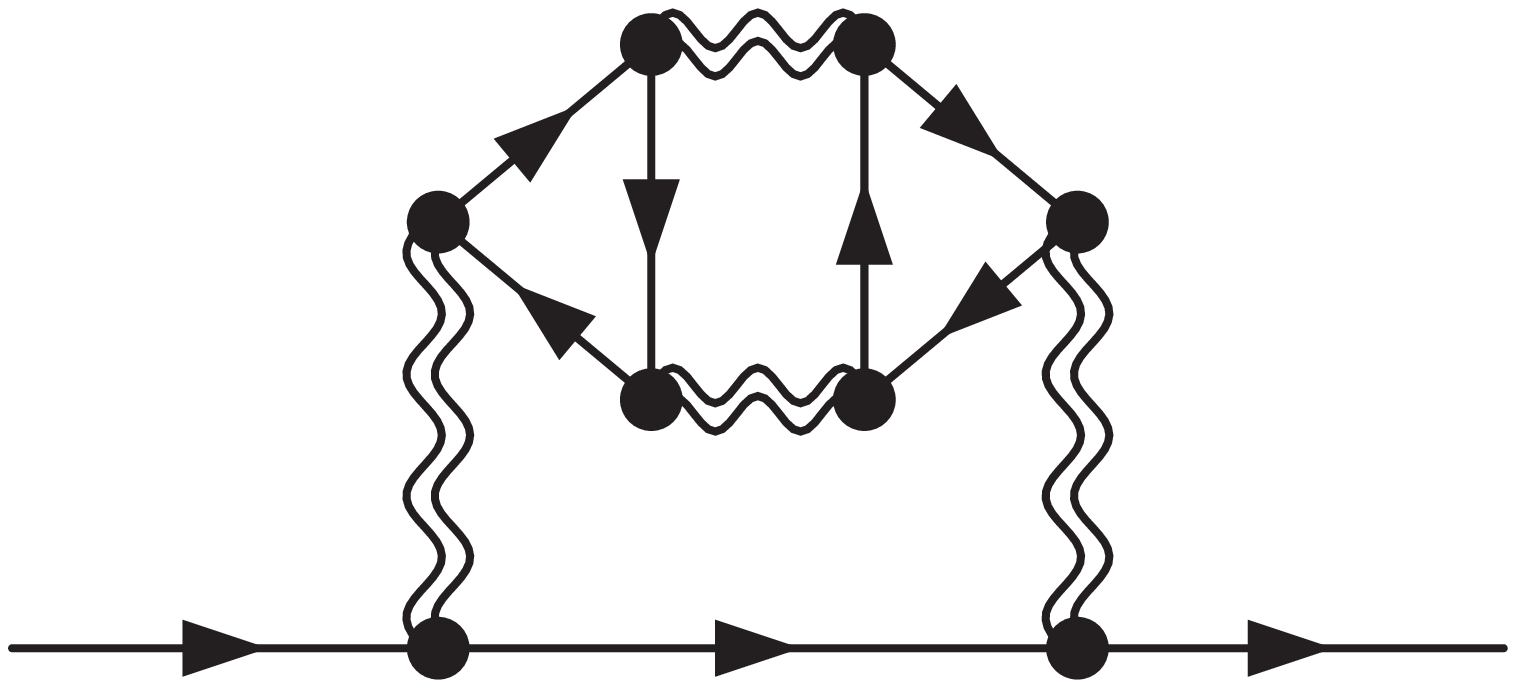}
 \rule{0.004\linewidth}{0cm}
 \includegraphics[width=0.24\linewidth]{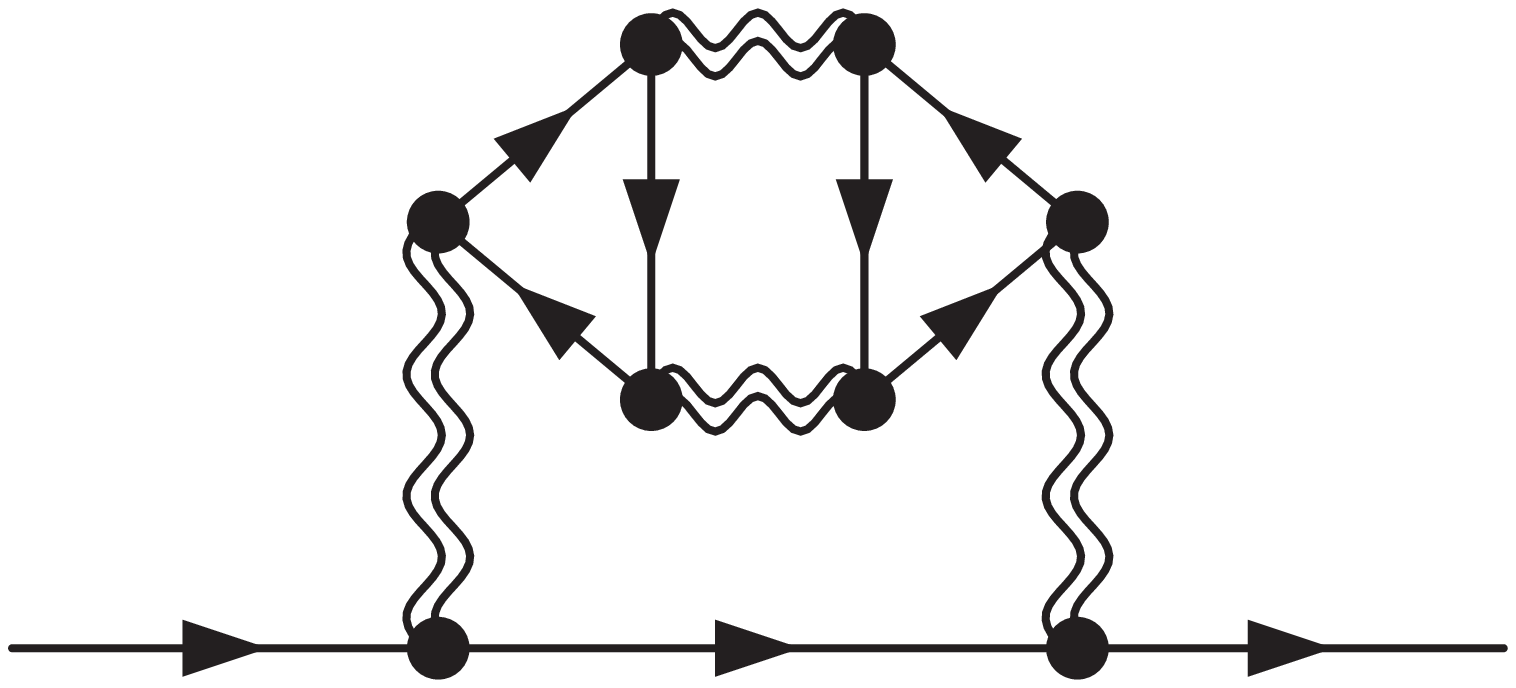}
 \end{minipage}
 \caption{
The order $1/(N-1)^2$ self-energy diagrams contribute away from half-filling  
in addition to the ones given in Fig.\ \ref{fig:sg_rpa_n2}.
The contribution of the particle-particle pairs 
and that of particle-hole pairs cancel each other 
in the particle-hole symmetric case. The counter term also vanishes, 
 $\lambda=0$, at half-filling.
}
\label{fig:sg_rpa_n2_asmX}
\end{figure}

We also consider 
the next-leading order term of the self-energy,
which we found in the previous work 
gives important contributions to the renormalization factor $z$ 
in the particle-hole symmetric case.\cite{AoSakanoFujii}
The order $1/(N-1)^2$  self-energy corrections 
arise from the diagrams shown in Figs.\ 
\ref{fig:sg_rpa_n2} and \ref{fig:sg_rpa_n2_asmX}.
Furthermore, the corrections arise also from 
the second and third diagrams in Fig.~\ref{fig:sg_rpa} 
through the higher-order component  
of $\mathcal{U}_\mathrm{bub}(i\omega)$ and 
that of $\lambda$, respectively.
In the particle-hole {\it asymmetric\/} case,  
 the diagrams indicated in Fig.\ \ref{fig:sg_rpa_n2_asmX} 
give finite contributions 
as the exact cancellation between the particle-particle 
and particle-hole pairs occurs only at half-filling.
We have calculated all these next-leading order contributions 
of the self-energy numerically 
to obtain $\widetilde{\gamma}$, $\lambda$, $E_d^*$ and 
the phase shift $\delta$ to order $1/(N-1)^2$.

%


\begin{figure}[t]
\leavevmode

\begin{minipage}[t]{0.8\linewidth}
\includegraphics[width=\linewidth]{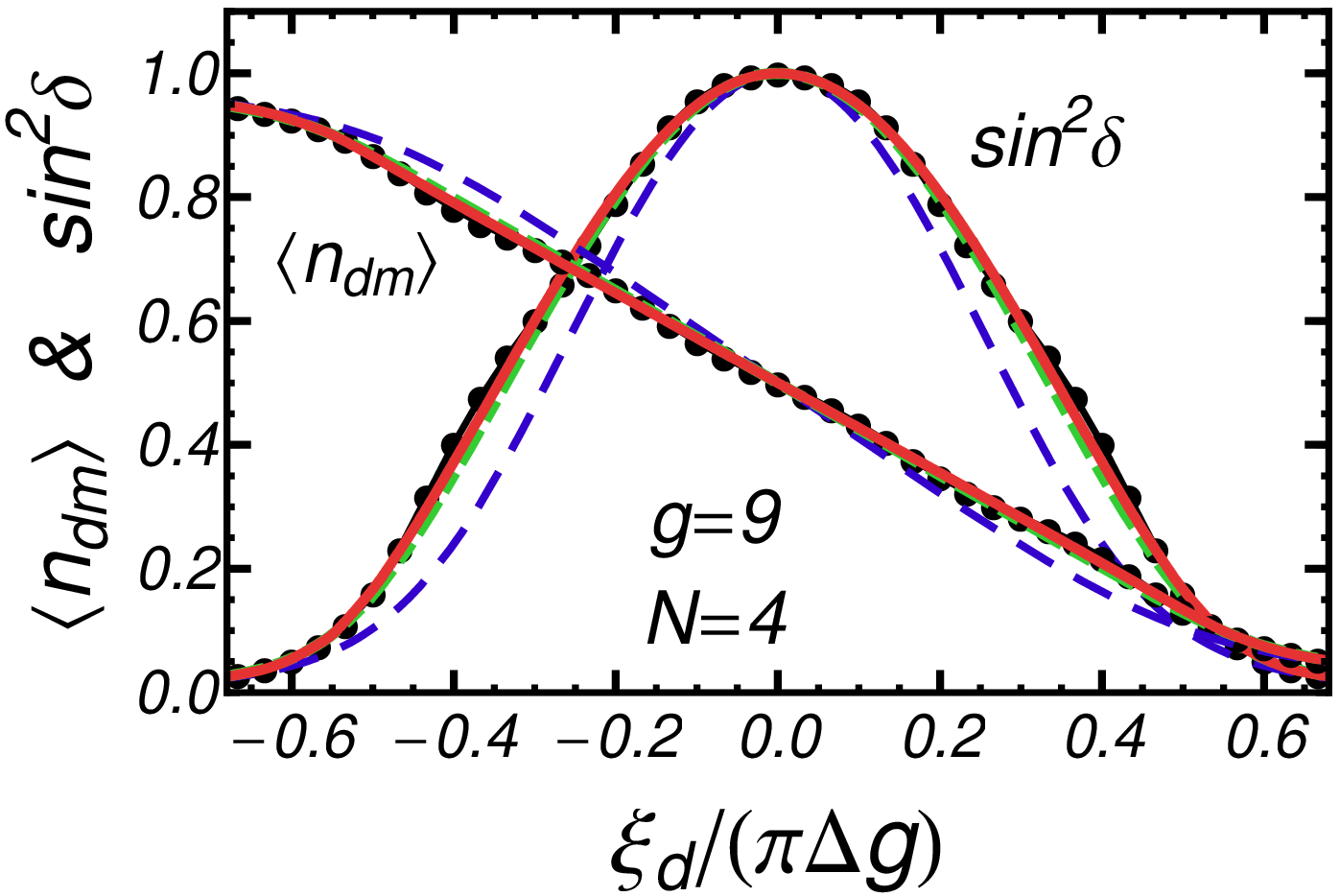}
\end{minipage}

\vspace{0.3cm}

\ \hspace{-0.5cm}
\begin{minipage}[t]{0.83\linewidth}
\includegraphics[width=\linewidth]{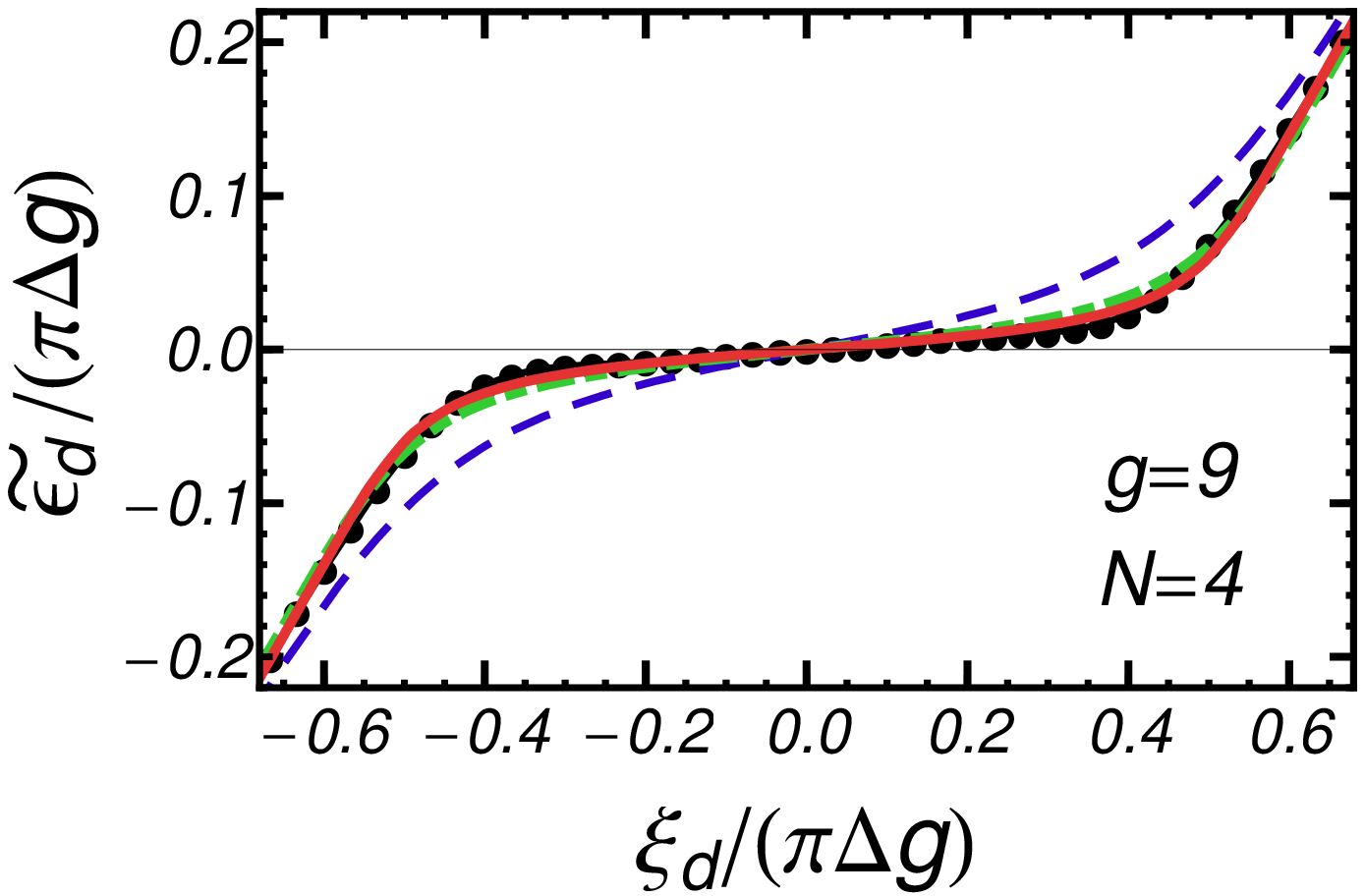}
\end{minipage}
\caption{
(Color online) 
$\langle n_{dm}\rangle=\delta/\pi$ and $\sin^2 \!\delta$ (upper panel), and 
$\widetilde{\epsilon}_d^{}$ (lower panel) plotted as 
a function of $\xi_d$ for $N=4$ and $g=9.0$.
The circles ($\bullet$) represent the NRG results,   
and the red solid line the order $1/(N-1)^2$ results.
The blue dashed line denotes the zeroth order results obtained 
with the HF approximation {\it without} the RPA corrections.
The order $1/(N-1)$ results, corresponding to the HF-RPA, 
are also plotted with the green dash-dot line 
although it is almost concealed under the red solid line.
}
\label{fig:nd_sin2d_eren}
\end{figure}


\begin{figure}[t]
\leavevmode

\ \hspace{-0.4cm}
\begin{minipage}[t]{0.82\linewidth}
\includegraphics[width=\linewidth]{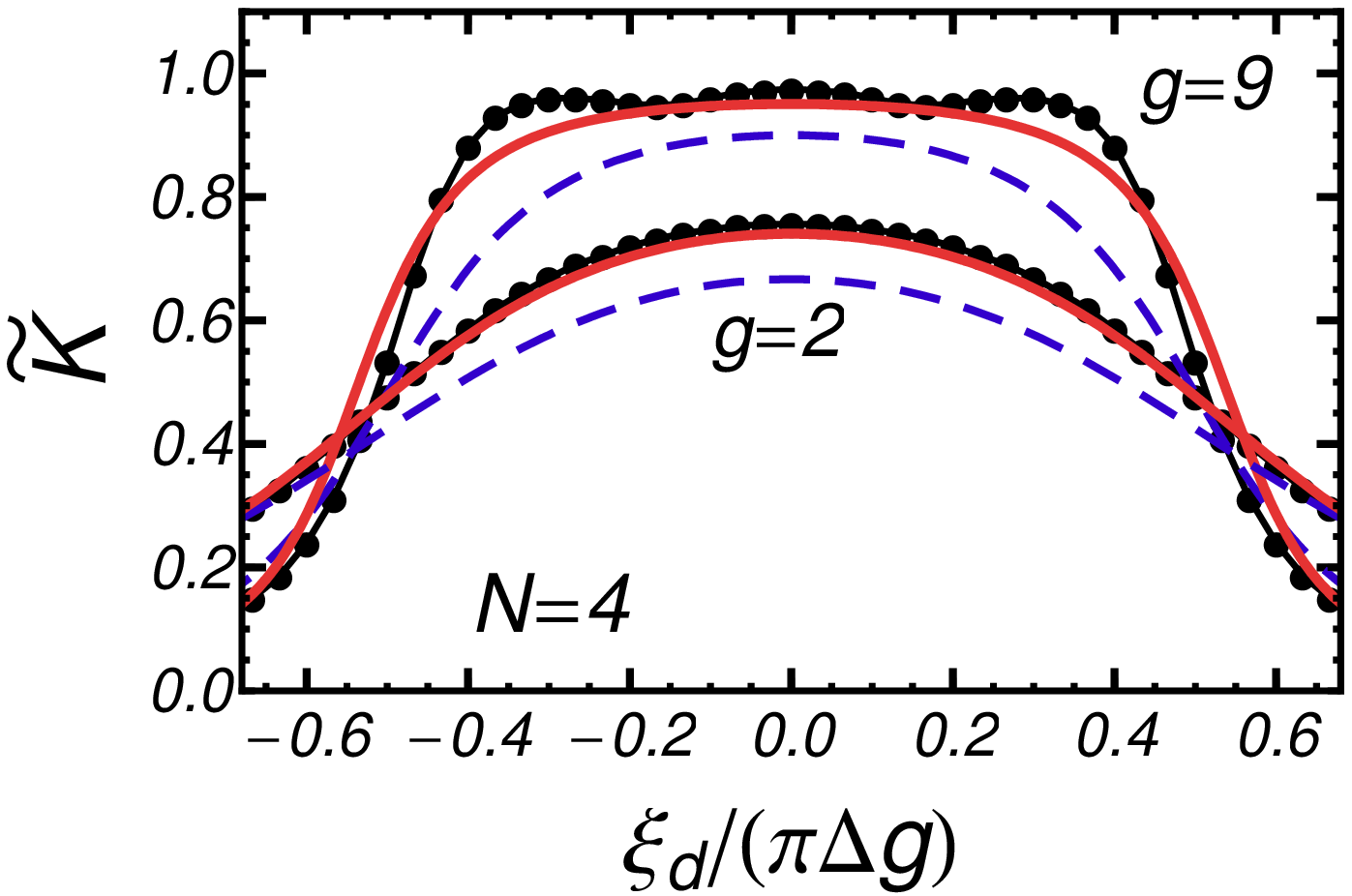}
\end{minipage}

\vspace{0.3cm}

\begin{minipage}[t]{0.8\linewidth}
\includegraphics[width=\linewidth]{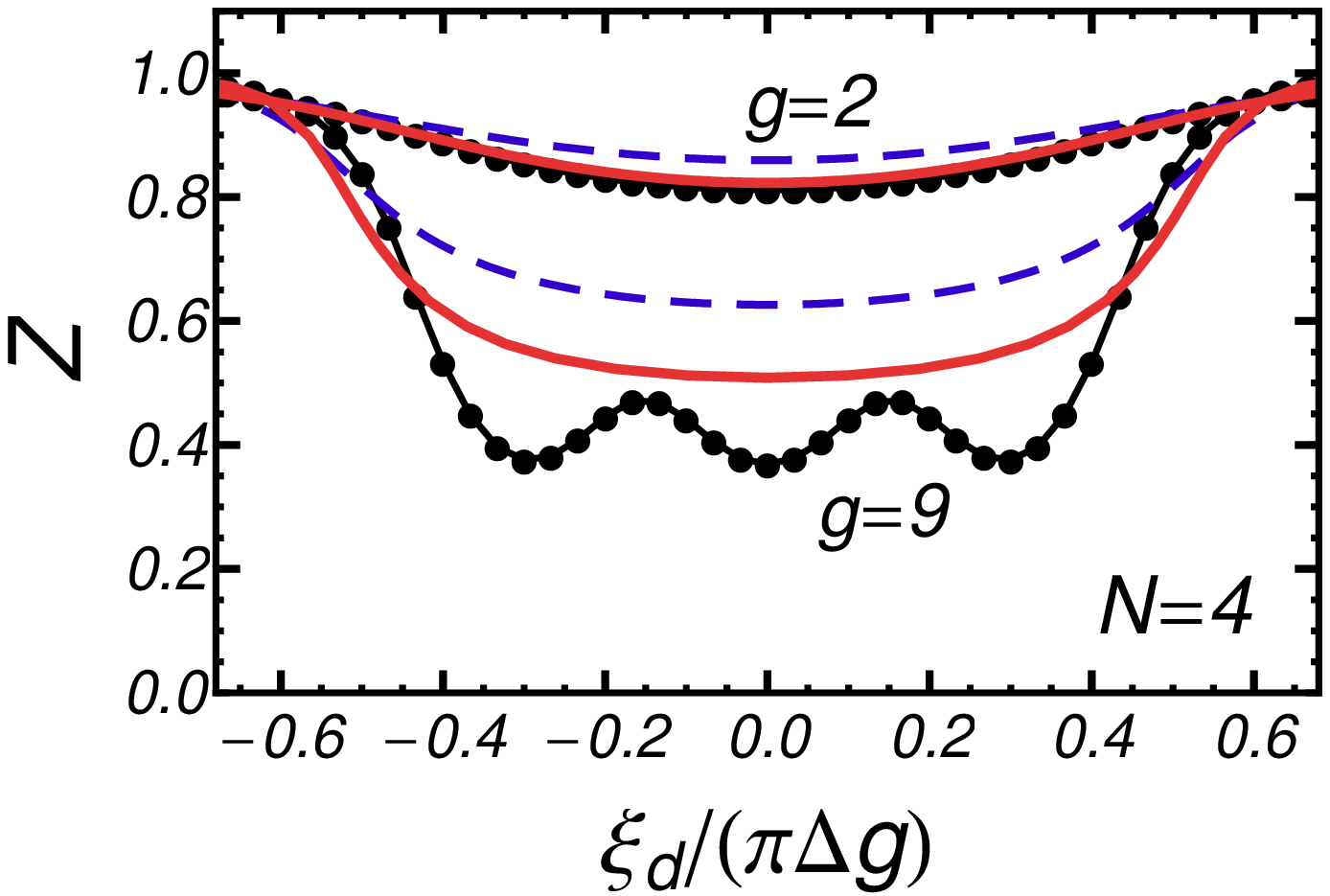}
\end{minipage}
\caption{
(Color online) 
Renormalized parameters  
$\widetilde{K}$ (upper panel) and $z$ (lower panel)
 plotted as a function of $\xi_d$ for $N=4$, choosing $g = 2.0$ and $9.0$.
The circles ($\bullet$) represent the NRG results.   
The blue dashed line represents 
the leading order results,  corresponding to the HF-RPA.
The read solid line represents the next-leading order results,
for which the fluctuation beyond the RPA are taken into account. 
}
\label{fig:Ktilde_Z}
\end{figure}
%



\begin{figure}[t]
\leavevmode

\ \hspace{-0.2cm}
\begin{minipage}[t]{0.8\linewidth}
\includegraphics[width=\linewidth]{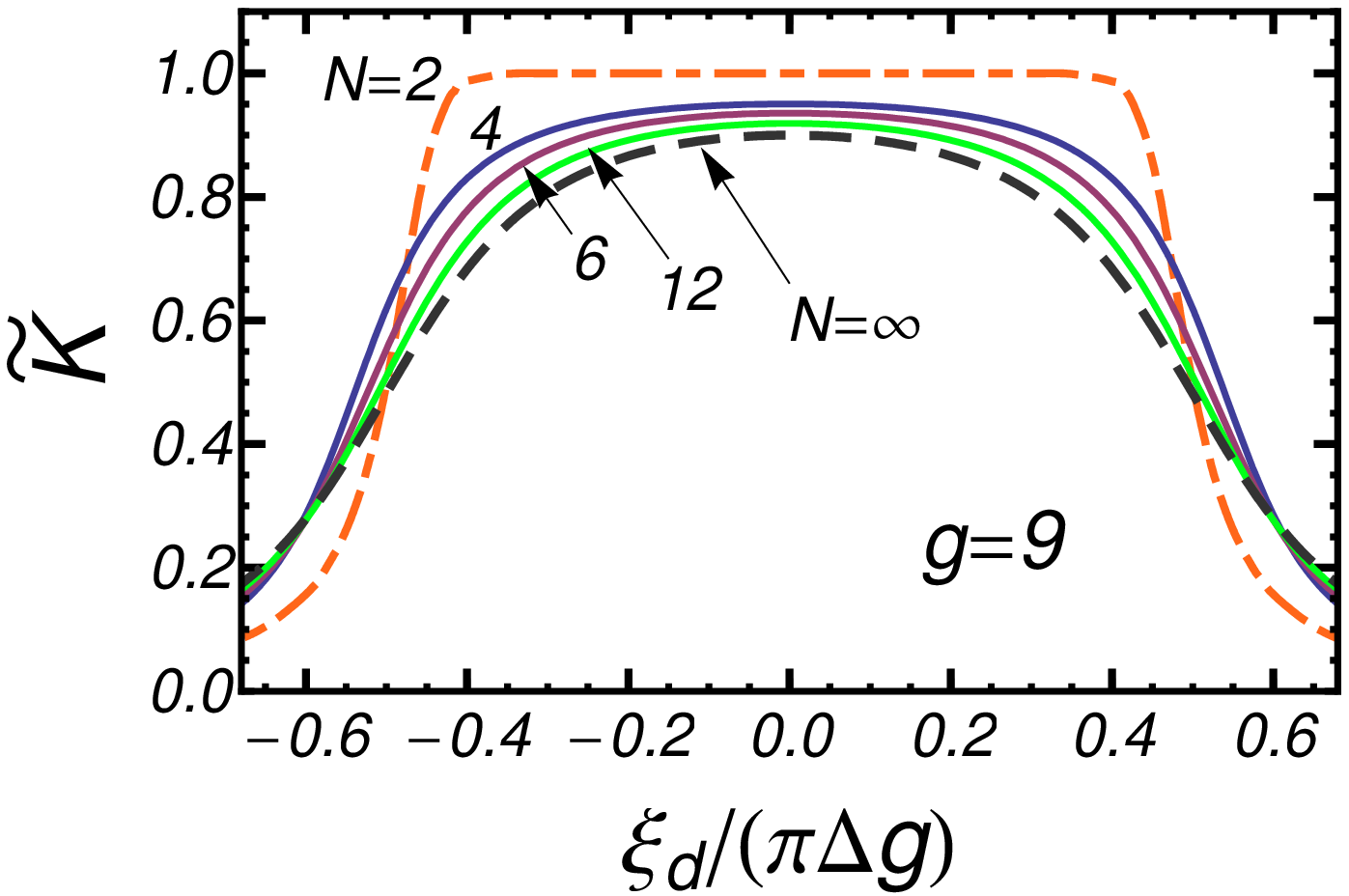}
\end{minipage}

\vspace{0.3cm}

\begin{minipage}[t]{0.8\linewidth}
\includegraphics[width=\linewidth]{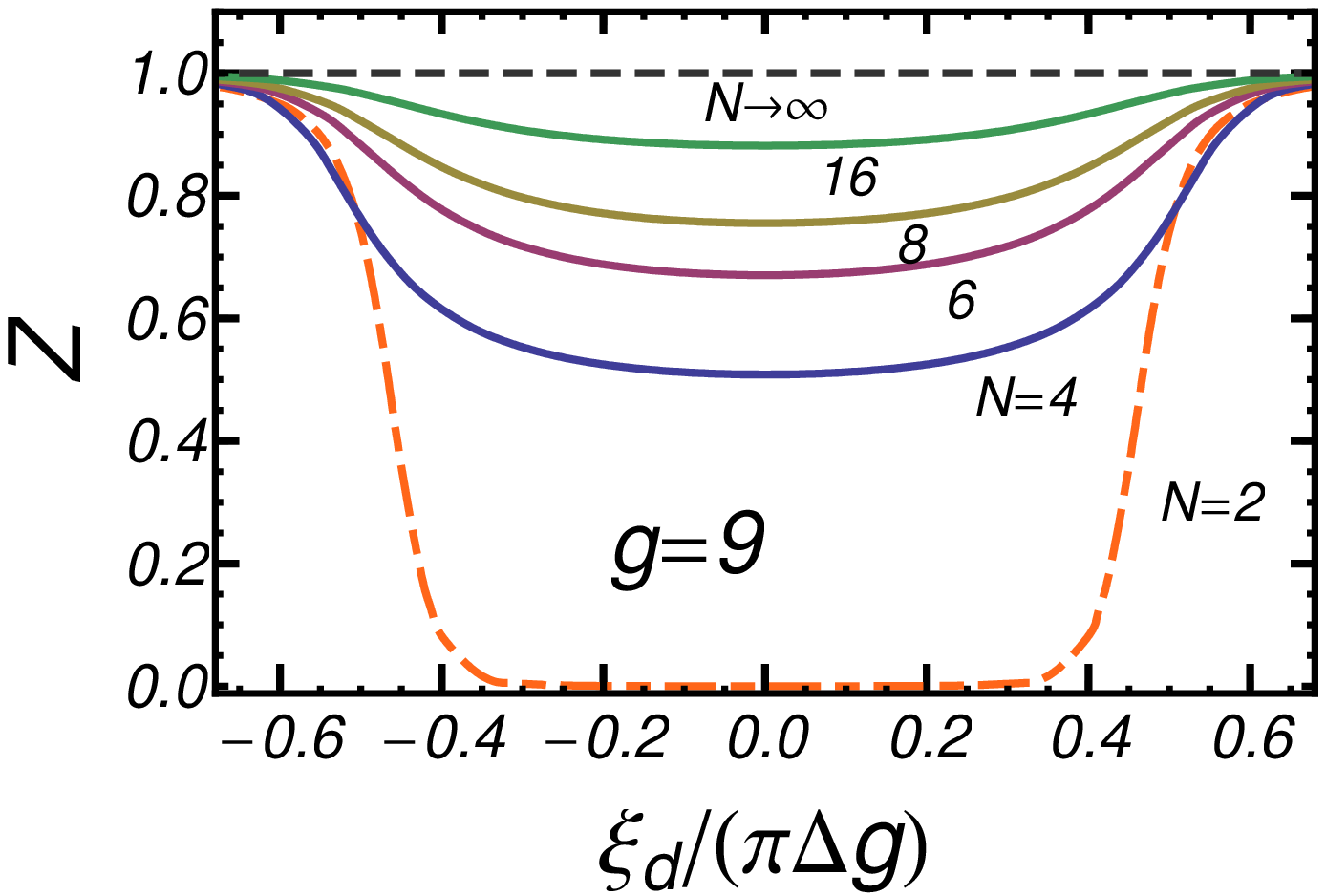}
\end{minipage}
\caption{
(Color online) 
Next leading-order results 
plotted vs $\xi_d$ for $g=9.0$,  
for several $N$ ($=4,6, \ldots, \infty$):
 $\widetilde{K}$ and $z$ are calculated 
up to order $1/(N-1)$ and $1/(N-1)^2$, respectively.
The orange dash-dot line represents the NRG results for $N=2$,
and the gray dashed line represents the exact results 
in the $N\to \infty$ limit.
}
\label{fig:N_dependenceG9}
\end{figure}

\begin{figure}[t]
\leavevmode

\ \hspace{-0.2cm}
\begin{minipage}[t]{0.8\linewidth}
\includegraphics[width=\linewidth]{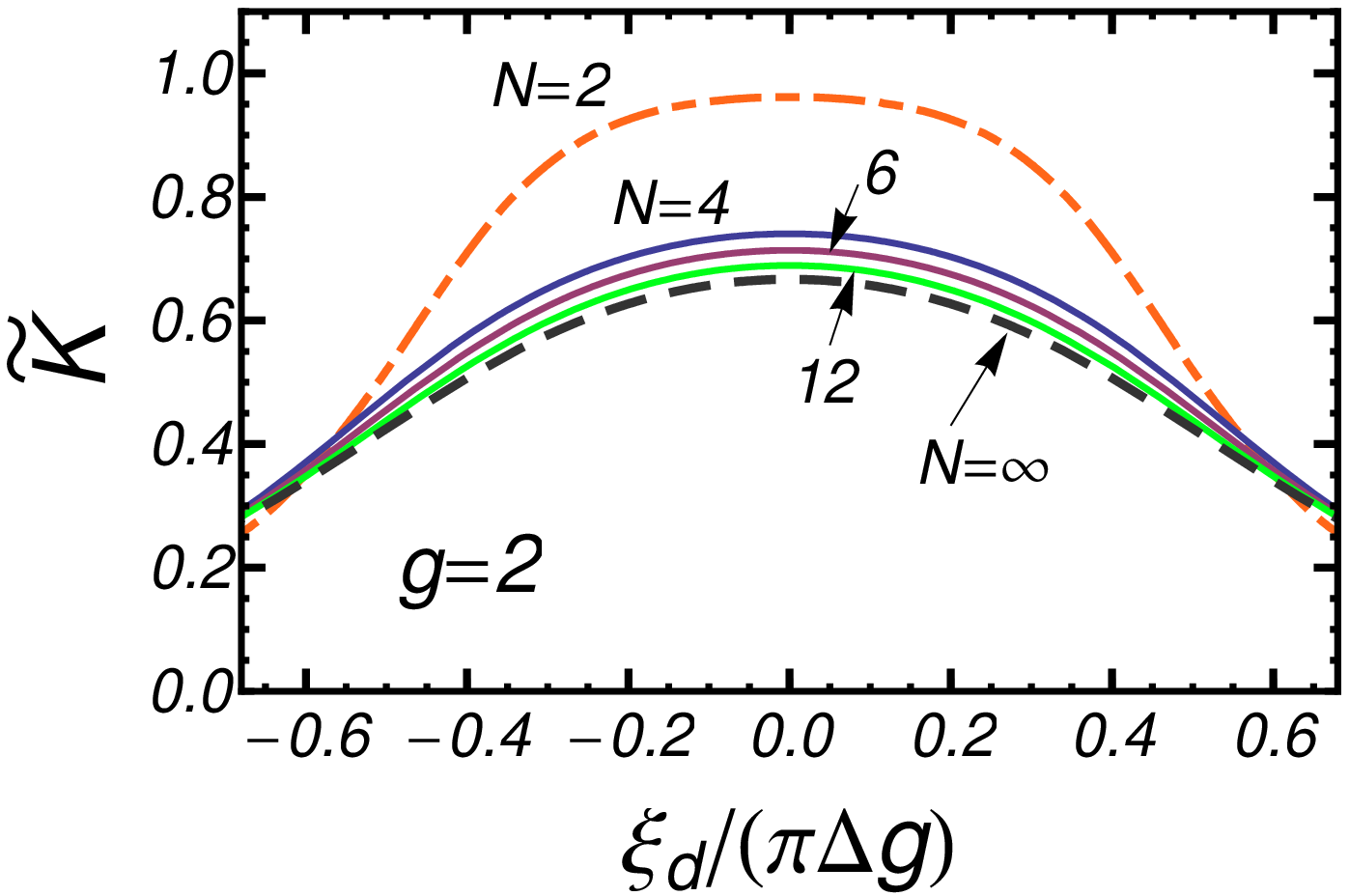}
\end{minipage}

\vspace{0.3cm}

\begin{minipage}[t]{0.8\linewidth}
\includegraphics[width=\linewidth]{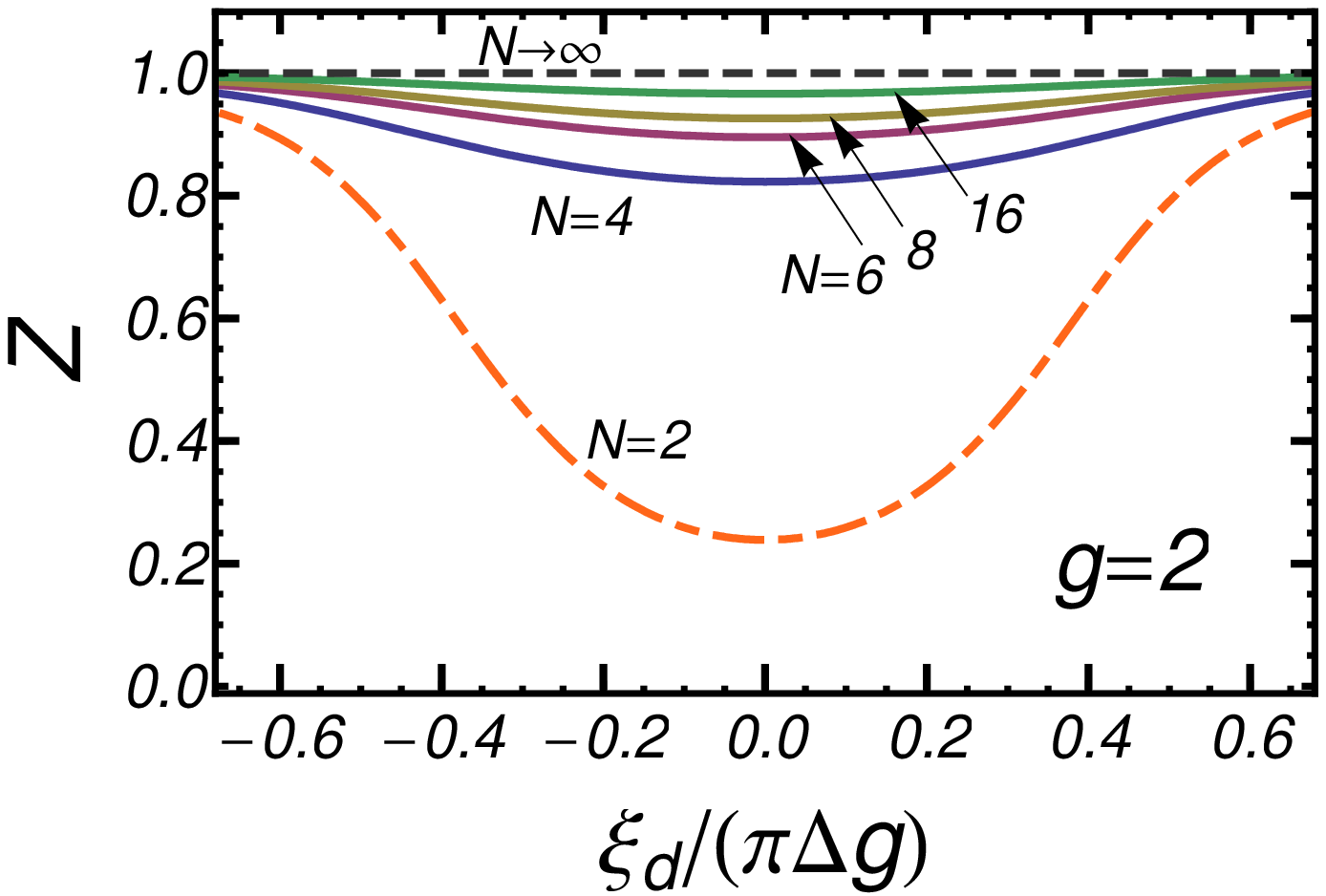}
\end{minipage}
\caption{
(Color online) 
Next leading-order results plotted vs $\xi_d$ for $g=2.0$, 
for several $N$ ($=4,6, \ldots, \infty$):  
 $\widetilde{K}$ and $z$ are calculated 
up to order $1/(N-1)$ and $1/(N-1)^2$, respectively.
The orange dash-dot line represents the NRG results for $N=2$,
and the gray dashed line represents the exact results 
in the $N\to \infty$ limit.
}
\label{fig:N_dependenceG2}
\end{figure}

\section{Numerical results}
\label{sec:results}

We have also carried out the NRG calculations for $N=4$, 
and have compared with the results of the $1/(N-1)$ expansion. 
We find generally that the next-leading order results 
agree well with the NRG results,
especially for small interactions $g \lesssim 3.0$. 
Typical examples are presented in 
Figs.\ \ref{fig:nd_sin2d_eren} and \ref{fig:Ktilde_Z}.

In Fig.\ \ref{fig:nd_sin2d_eren},  
comparisons are made for $\langle n_{dm} \rangle$, $\sin^2 \! \delta$ 
and $\widetilde{\epsilon}_d$, 
choosing a rather large value $g=9.0$ for the interaction  
in order to show clearly the small deviations.
We see the very close agreement between 
the order $1/(N-1)^2$ results (red sold line) and
 the NRG ($\bullet$)  results over the whole region of $\xi_d$.
The order $1/(N-1)$ results, corresponding to the HF-RPA,  
are also plotted in Fig.\ \ref{fig:nd_sin2d_eren} 
with the green dash-dot line  
although they are almost concealed under the red solid line. 
Nevertheless, a very small deviation is visible 
near $\xi_d/(\pi \Delta g) \simeq \pm 0.4$, 
showing that the order $1/(N-1)^2$ corrections improve 
the results slightly closer to the exact NRG results. 
It indicates that the phase shift $\delta$ can be approximated  
reasonably by the order $1/(N-1)$ self-energy 
given in Eq.\ \eqref{eq:sg_to_order2_w0}, 
which significantly improves the simple HF values (blue dashed line)
given at the zeroth order {\it without\/} 
taking into account the RPA fluctuations.
We have also confirmed similar trend 
for a larger value of the repulsion $g=12.0$.
We also see in the lower panel of Fig.\ \ref{fig:nd_sin2d_eren} that 
the renormalized impurity level  
$\widetilde{\epsilon}_d$ stays close to the Fermi level   
in the region $|\xi_d|/(\pi \Delta g) \lesssim 0.3$ 
as the interaction $g$ is relatively large in this case. 
Outside of this region 
the impurity states tend to be empty at $\xi_d/(\pi \Delta g) \gtrsim 0.5$, 
or fully occupied at $\xi_d/(\pi \Delta g) \lesssim -0.5$.
We note that the NRG results 
for $\sin^2 \delta$ and $\langle n_{dm} \rangle$  
wave weakly at $\xi_d/(\pi \Delta g) \simeq \pm 0.3$, 
relating to the Coulomb oscillation discussed below.

In Fig.\ \ref{fig:Ktilde_Z}, 
the renormalized parameters 
$\widetilde{K}$ and $z$ are plotted for $N=4$, 
choosing $g=2.0$ and $9.0$.
We see that the next-leading order corrections (red solid line) 
 significantly improve the leading order results   
corresponding to the HF-RPA (blue dashed line), 
towards the exact NRG values (solid circles).  
Specifically, for small interactions ($g=2.0$), 
the close agreement between the next-leading order results 
and that of the NRG are seen both for $\widetilde{K}$ and $z$.
However, for large interactions ($g=9.0$), 
the NRG results show an oscillatory behavior,  
as reported recently also in Ref.\ \onlinecite{Nishikawa2}.
This behavior is seen more clearly for the renormalization factor $z$ 
than the coupling $\widetilde{K}$.
The oscillation is caused by addition of electron into the local orbitals, 
which in the atomic limit of $\Delta =0$ occurs   
discontinuously at $-\epsilon_d/U =0,1,2,\ldots,N-1$. 
Thus, for strong Coulomb interactions, 
$N-1$ local minima (maxima) appear 
for $z$ ($\widetilde{K}$) 
in the region of $|\xi_d|/(\pi \Delta g) \leq 0.5$, 
which corresponds to  $-(N-1)U \leq \epsilon_d \leq 0$.

Truncation of the $1/(N-1)$ expansion 
does not reproduce the oscillatory behavior. 
Nevertheless, we see in Fig.\ \ref{fig:Ktilde_Z} 
that the next-leading order results of  
 $\widetilde{K}$ are still close to the exact NRG results 
even for large $g=9.0$, 
especially at the plateau region for 
$|\xi_d|/(\pi \Delta g) \lesssim 0.2$,  
where $\widetilde{K}$ reaches near the unitary-limit value $1.0$ 
and the charge excitation is suppressed significantly.
Furthermore, we also see that the order $1/(N-1)^2$ results of 
 $z$ for $g=9.0$ approach closely the maxima, 
emerging  
at $\xi_d/(\pi \Delta g) \simeq \pm 0.17$ in the NRG results,
which correspond to mixed-valence regions 
in between two adjacent integer charge states. 
Therefore,  the fluctuations beyond the RPA,  
taken into account to order $1/(N-1)^2$, 
can describe approximately 
an upper envelop for the waving curve 
of the renormalization factor, $z$,  for $N \geq 4$.
Note that  the Kondo energy scale can also be 
estimated from this result of $z$ 
as the width of the Kondo resonance 
is given by $\widetilde{\Delta} = z \Delta$.

In order to clarify the dependence of the 
renormalized parameters on the orbital degeneracy $N$, 
the next-leading order results of $\widetilde{K}$ and $z$ 
for several values of $N$ ($=4,\,6, \ldots$) are shown  
in Figs.\ \ref{fig:N_dependenceG9} and \ref{fig:N_dependenceG2} 
for $g=9.0$ and $2.0$. 
In these figures, the NRG results for $N=2$ (orange dash-dot line)  
and the exact values in the $N \to \infty$ limit 
(gray dashed line) are also shown.
We see, in both strong $g=9.0$ and weak $g=2.0$ coupling cases, 
the value that the renormalized coupling $\widetilde{K}$ 
can take at $N \geq 4$ is bounded  
in a narrow region between the curve for $N=4$ 
and the one for the large $N$ limit that is 
described by the RPA given in Eq.\ \eqref{eq:g_inf}.
As $N$ increases,  $\widetilde{K}$ converges rapidly 
to the asymptotic value in the $N \to \infty$ limit.

The zeroth-order HF results 
for $\langle n_{dm} \rangle$, $\sin^2\! \delta$ 
and $\widetilde{\epsilon}_d$, 
which are plotted in Fig.\ \ref{fig:nd_sin2d_eren} with 
the blue dashed lines, 
also correspond to the exact values in the large $N$ limit.
Therefore, as $N$ increases from $N=4$ to $\infty$,
the phase shift $\delta$ and the renormalized impurity level 
$\widetilde{\epsilon}_d$ vary also 
in a narrow region between the red-solid and blue-dashed lines 
indicated in Fig.\ \ref{fig:nd_sin2d_eren}.

In contrast, we see in the lower panel 
of Figs.\ \ref{fig:N_dependenceG9} and \ref{fig:N_dependenceG2}
that the renormalization factor  $z$ varies in a wider range.
Specifically, for large interaction $g=9.0$, 
it varies in the range of $ 0.4\lesssim z \leq 1.0$ 
at $|\xi_d|/(\pi \Delta g) \lesssim 0.3$ 
as the orbitals increases from $N=4$ to $N \to \infty$. 
As mentioned for Fig.\ \ref{fig:Ktilde_Z}, 
the next-leading order results of $z$ become  
less accurate than that of $\widetilde{K}$ for large $g$. 
Nevertheless, we also see in Fig.\ \ref{fig:Ktilde_Z}  
that for small interactions ($g=2.0$) 
the order $1/(N-1)^2$ results of $z$ are very close 
to the exact NRG values even for $N=4$ 
where the degeneracy $N$ is not so large. 
Furthermore, as $N$ increases,  
the order $1/(N-1)^2$ corrections dominate the fluctuations 
beyond the RPA, and capture reasonably the correlation effects, 
especially in the mixed-valence regime.

\section{Summary}
\label{sec:summary}

In summary, 
we have described the $1/(N-1)$ expansion approach  
to the SU($N$) Anderson model with finite Coulomb interaction,  
without assuming the particle-hole symmetry. 
We have calculated the renormalized parameters
up to order $1/(N-1)^2$.
Specifically, the results of the phase shift $\delta$, or $E_d^*$, 
show very close agreement with the NRG results for $N=4$. 
This trend has been confirmed for $g \lesssim 10.0$, 
from small to relatively large interactions. 
The scaled Wilson ratio $\widetilde{K}=(N-1)(R-1)$ and the 
renormalization factor $z$ also agree well with the exact value 
in the mixed-valence regime, or not too large interactions.
As $N$ increases from $N=4$, 
$\widetilde{K}$ varies in a narrow range and converges 
rapidly to the RPA value that is asymptotically exact 
in the large $N$ limit,  
whereas $z$ varies in a wider range.
 This behavior of $\widetilde{K}$ and $z$ 
is determined essentially by the fluctuations beyond the HF-RPA, 
captured through the order $1/(N-1)^2$ corrections.

The $1/(N-1)$ expansion provides a well-defined and controlled way 
to take into account the fluctuations beyond the HF-RPA.
Furthermore, the expansion scheme based on  
the standard Feynman diagrammatic formalism for fermions with two-body 
interactions is quite general and flexible.
Therefore, this approach has wide potential applications, 
such as the nonequilibrium problem in the Keldysh formalism  
and the Hubbard-type lattice models.

\begin{acknowledgments}
We thank R. Sakano, T.\ Fujii, and A.\ C.\ Hewson 
for helpful discussions.
This work is supported by 
a JSPS Grant-in-Aid for 
Scientific Research C (No.\ 23540375).
Numerical computation was partly carried out 
at Yukawa Institute Computer Facility.
\end{acknowledgments}


\appendix

\section{Coefficients for the bubble diagrams}
\label{sec:bubble_coefficient}

The contribution of the bubble diagram 
on the vertex correction for $m \neq m'$,
shown in Fig.\  \ref{fig:vertex_rings}, 
is {\it not\/} a simple geometric series in powers of $U$, 
\begin{align}
&
\!\!\!\! 
\mathcal{U}_\mathrm{bub}(i\omega) 
\, =  \, 
U+  U\sum_{k=1}^{\infty} \, 
 \mathcal{A}_k\,\Bigl[- U\,\chi_0(i\omega) \Bigr]^k \;,  
\label{eq:U_bub_def}
\end{align}
where $\chi_0(i\omega)$ represents the contribution of a single loop,
\begin{align}
 \!\!\!\!
 \chi_0^{}(i\omega)  \,\equiv &\,   
  - \int_{-\infty}^{\infty}
  \! \frac{d\omega'}{2\pi} \,
 G_0^{}(i\omega +i\omega') \,G_0^{}(i\omega')
 \nonumber \\
  =& \,   \frac{1}{\pi}\, 
  \frac{\Delta}{|\omega|(|\omega|+2\Delta)}  \,
 \log\! \left[\frac{\left(|\omega|+\Delta\right)^2 + {E_d^*}^2}
 {\Delta^2+ {E_d^*}^2}\right] . 
 \end{align}
Specifically, 
$\pi \Delta \chi{_0^{}}(0) =  1/[1+(E_d^*/\Delta)^2]$ 
in the static limit $\omega \to 0$.
The coefficient $\mathcal{A}_k$ in Eq.\ \eqref{eq:U_bub_def}  
arises from the summations over the orbital indices 
for the series of $k$ fermion loops, 
which can be calculated as 
\begin{align}
\mathcal{A}_k \,=&\ 
{\sum_{m_1}}' {\sum_{m_2}}'\cdots {\sum_{m_k}}'\, 1 
\ =\, 
\frac{(N-1)^{k+1}-(-1)^{k+1}}{N}
\nonumber \\
=& \ (N-1)^k \, \sum_{p=0}^k \, \left(\frac{-1}{N-1} \right)^p
\;.
\label{eq:coefficient_bubble}
\end{align}
Here,  the primed sums for $m_j$'s run under 
the restriction due to the Pauli exclusion; 
 $m_j \neq m_{j+1}$ for $j=0,1,\ldots, k$ with 
$m_0=m$ and $m_{k+1} =m'$. 
Equation \eqref{eq:coefficient_bubble} shows that 
the coefficient $\mathcal{A}_k$ can be regarded 
as a polynomial of $(N-1)$,
and  $\mathcal{U}_\mathrm{bub}(i\omega)$ has   
the higher-order components in the expansion 
with respect to $1/(N-1)$.


\end{document}